\documentstyle[preprint,eqsecnum,aps,tighten]{revtex}

\begin{document}
\draft
\title{$E_{1g}$ model of superconducting UPt$_3$}
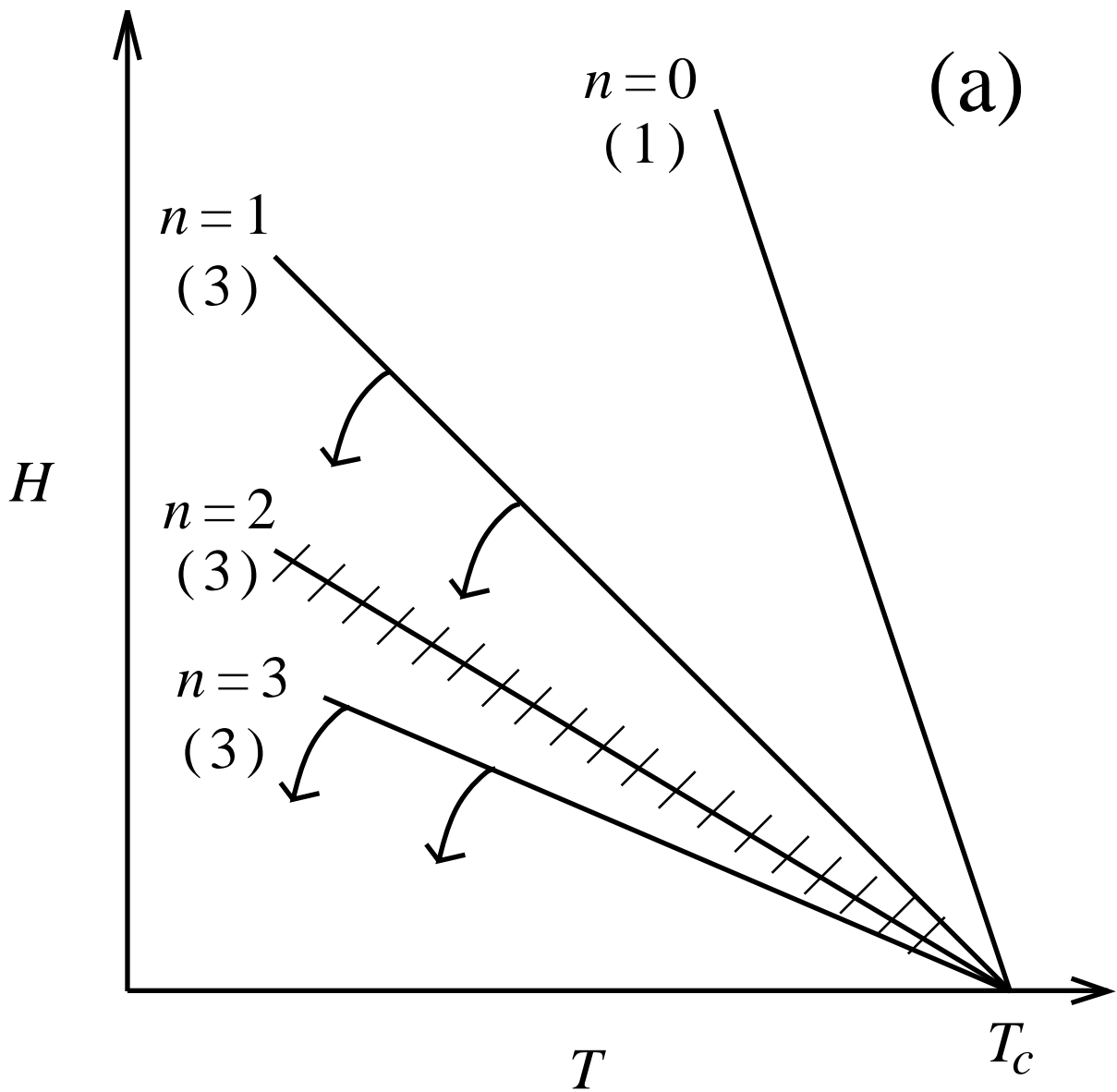
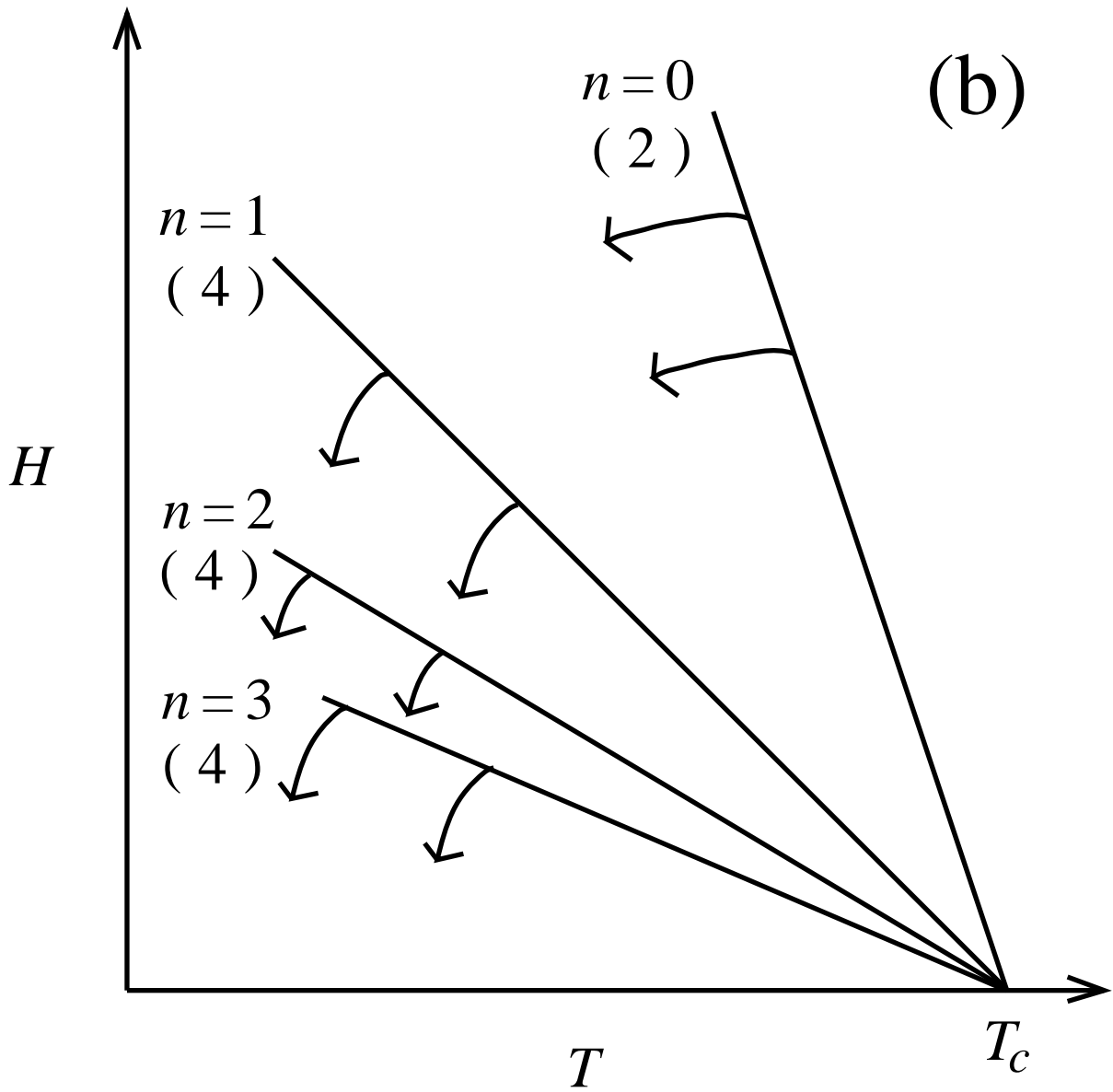
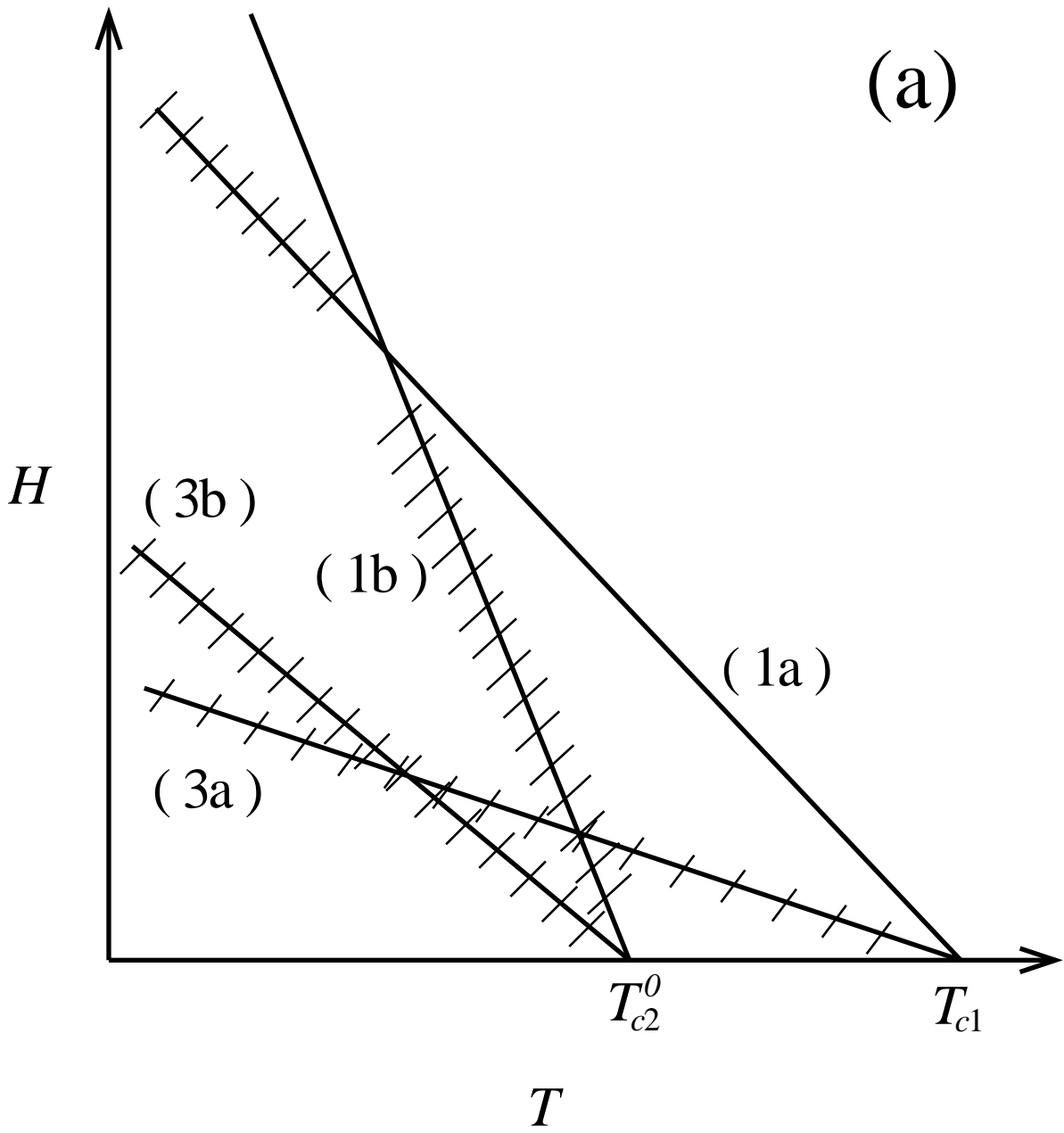
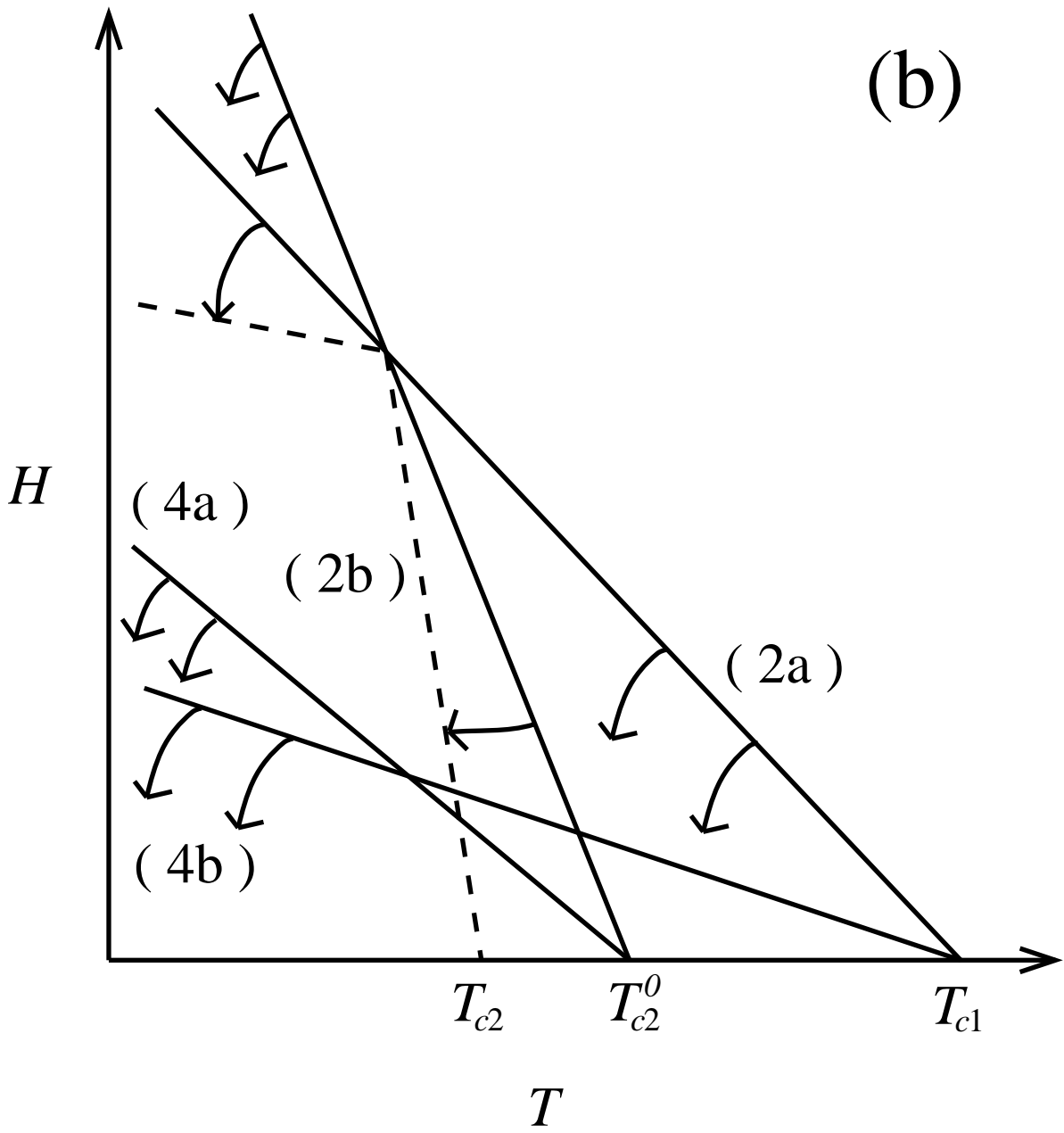
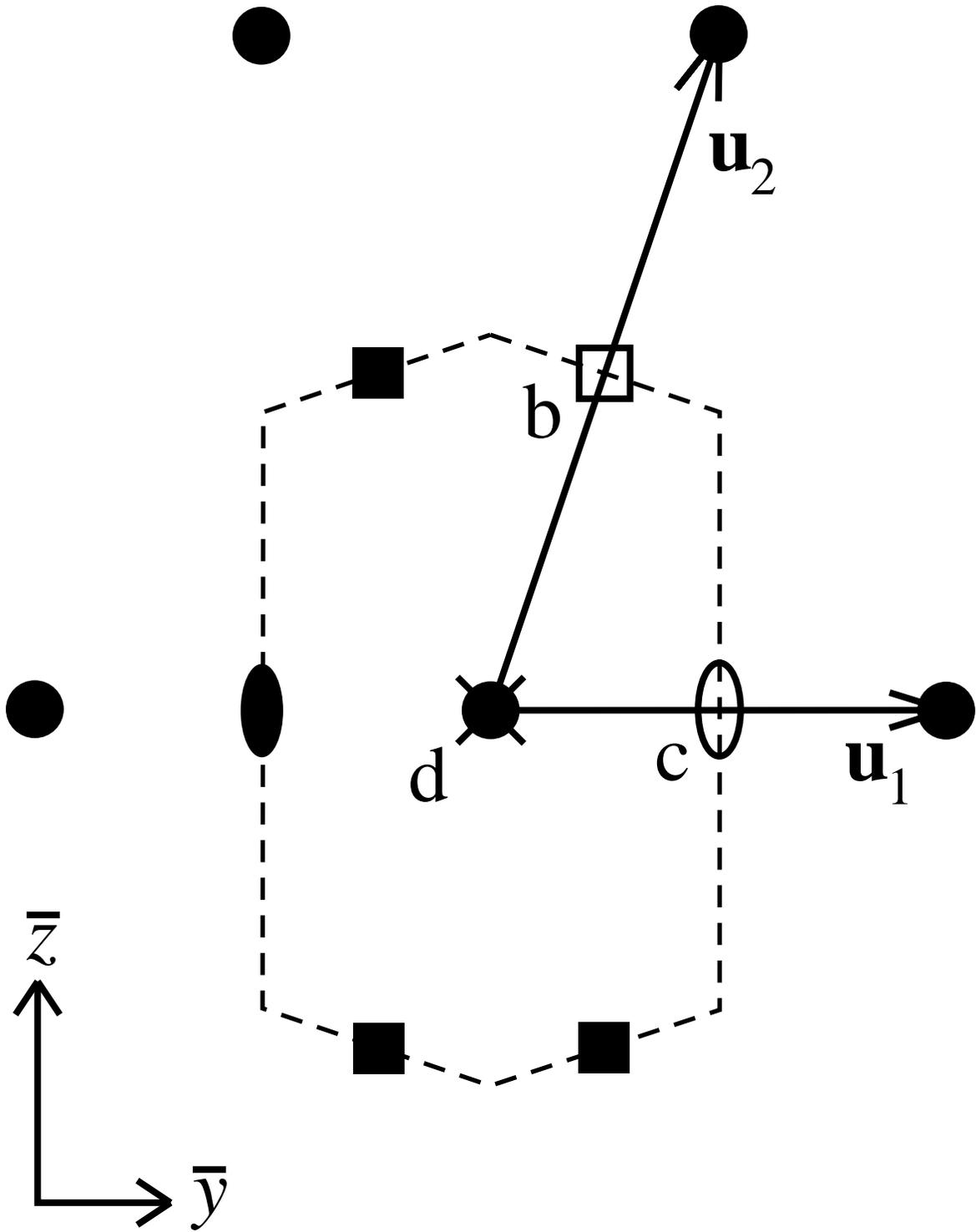
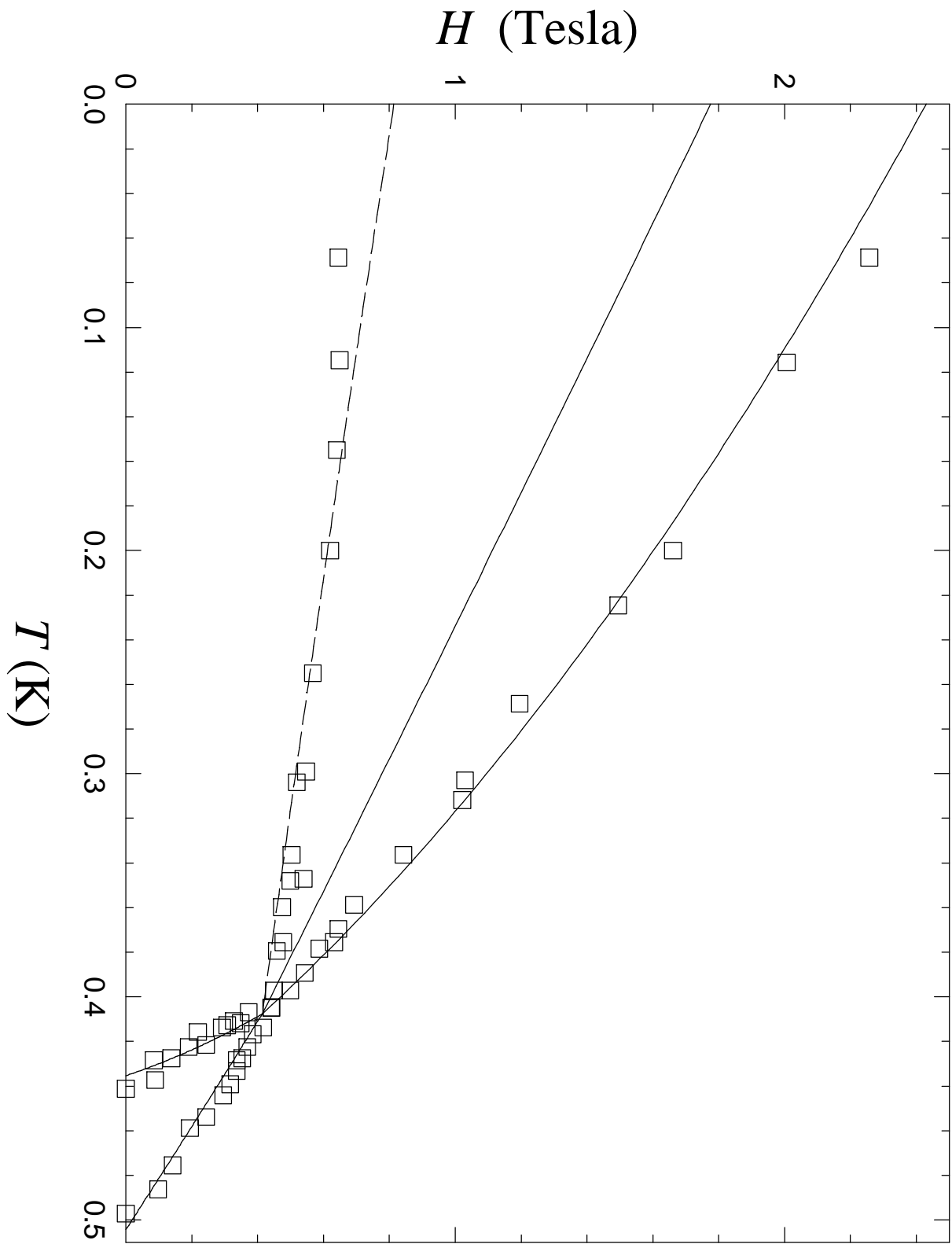
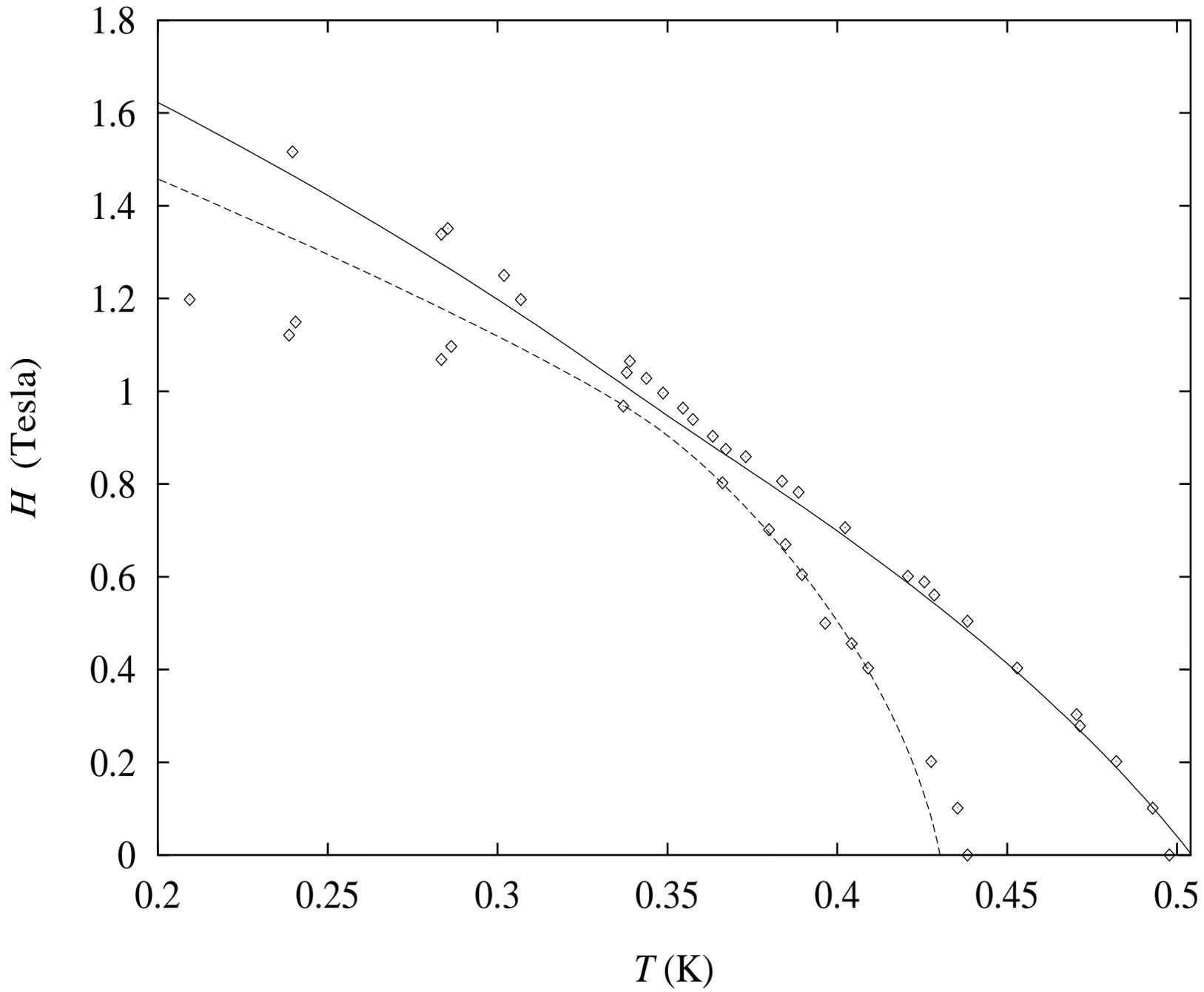
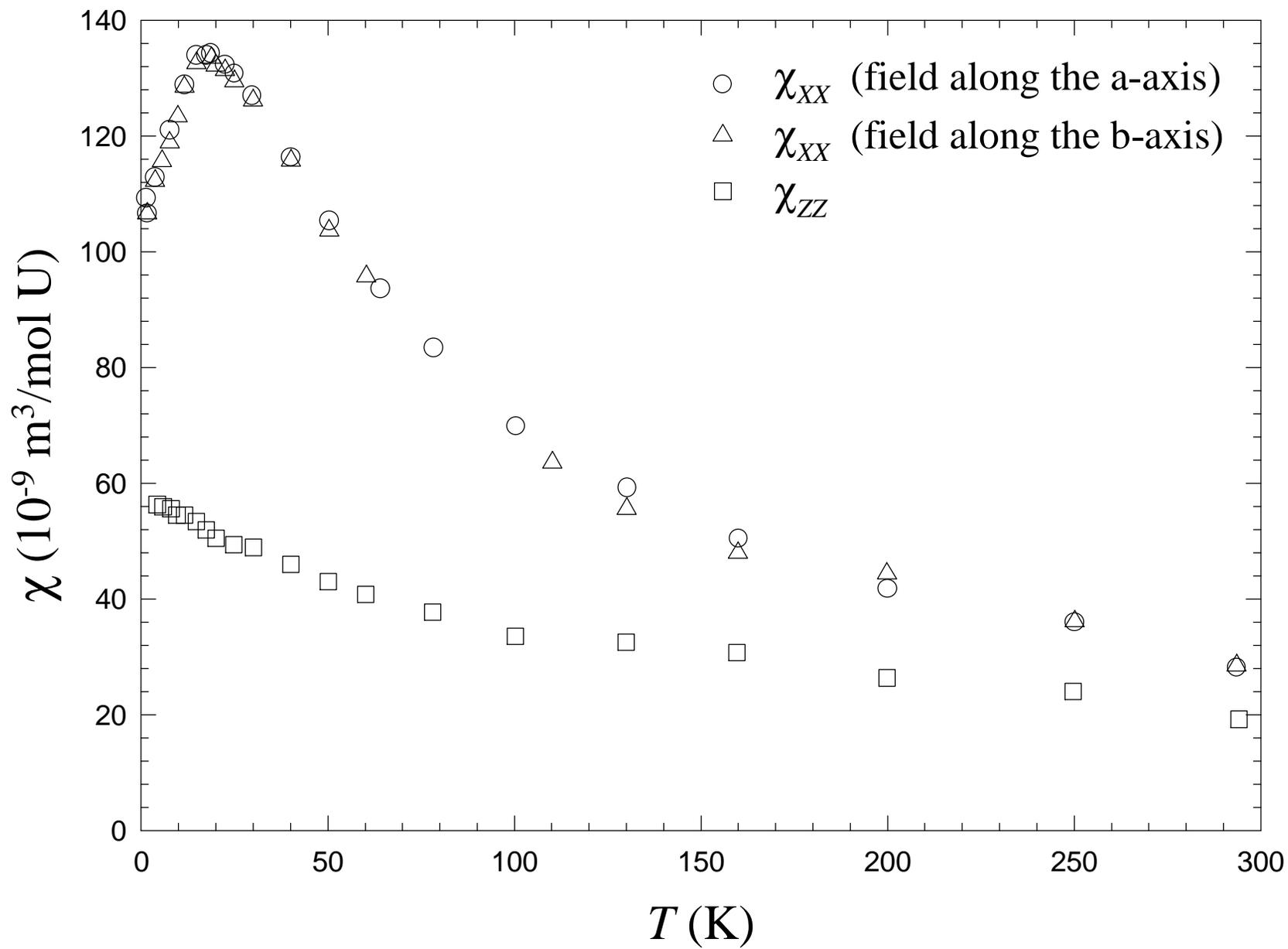
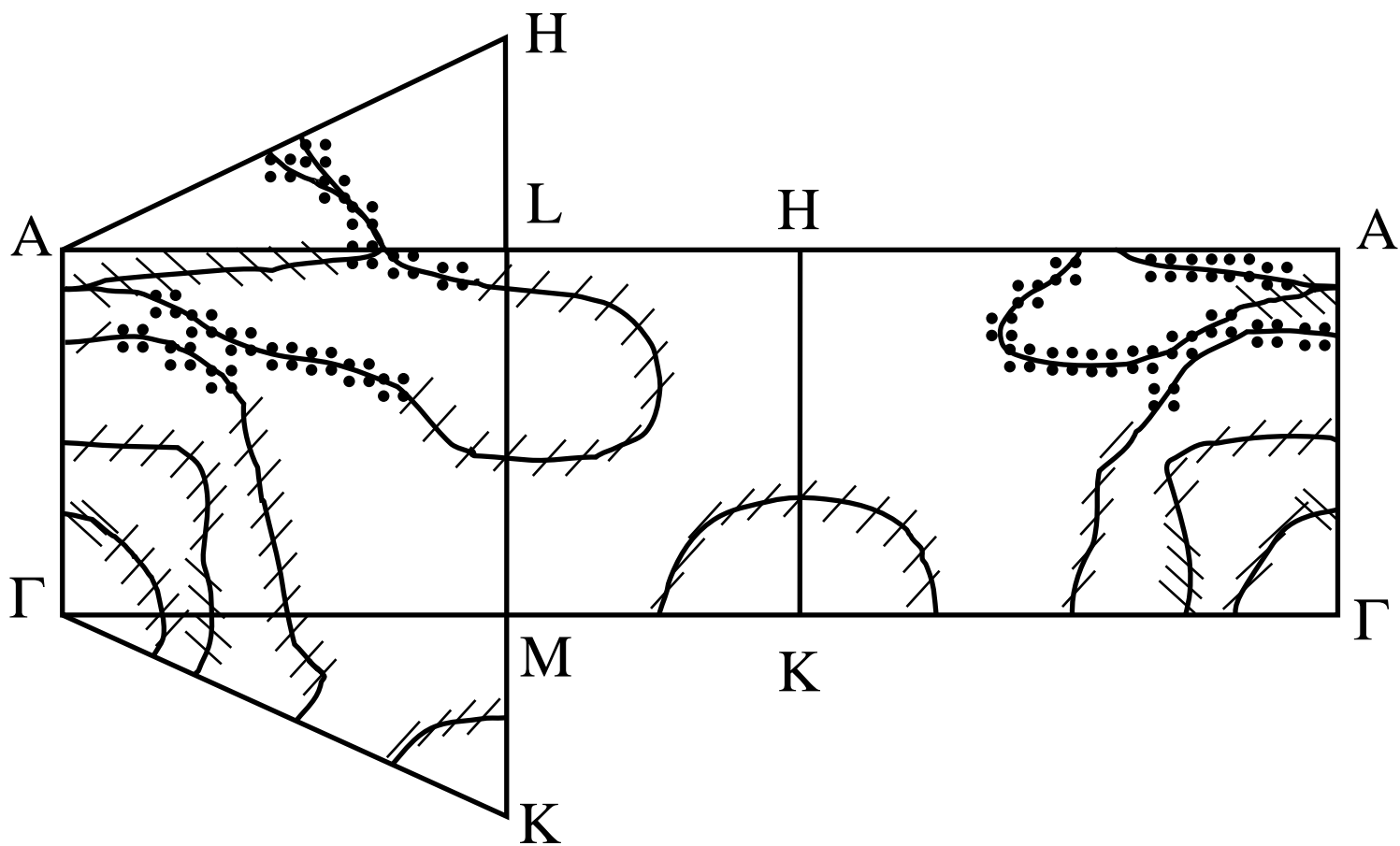
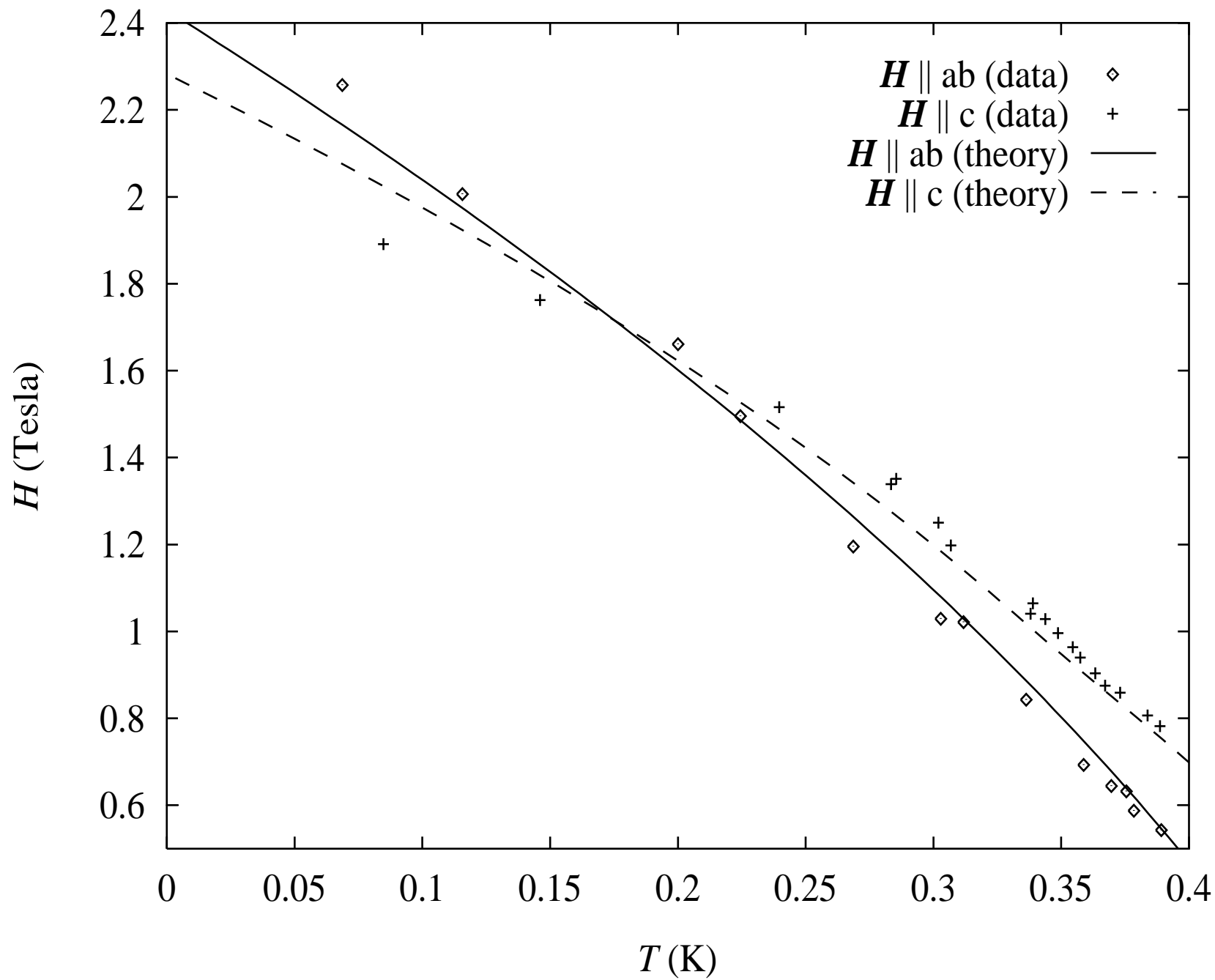
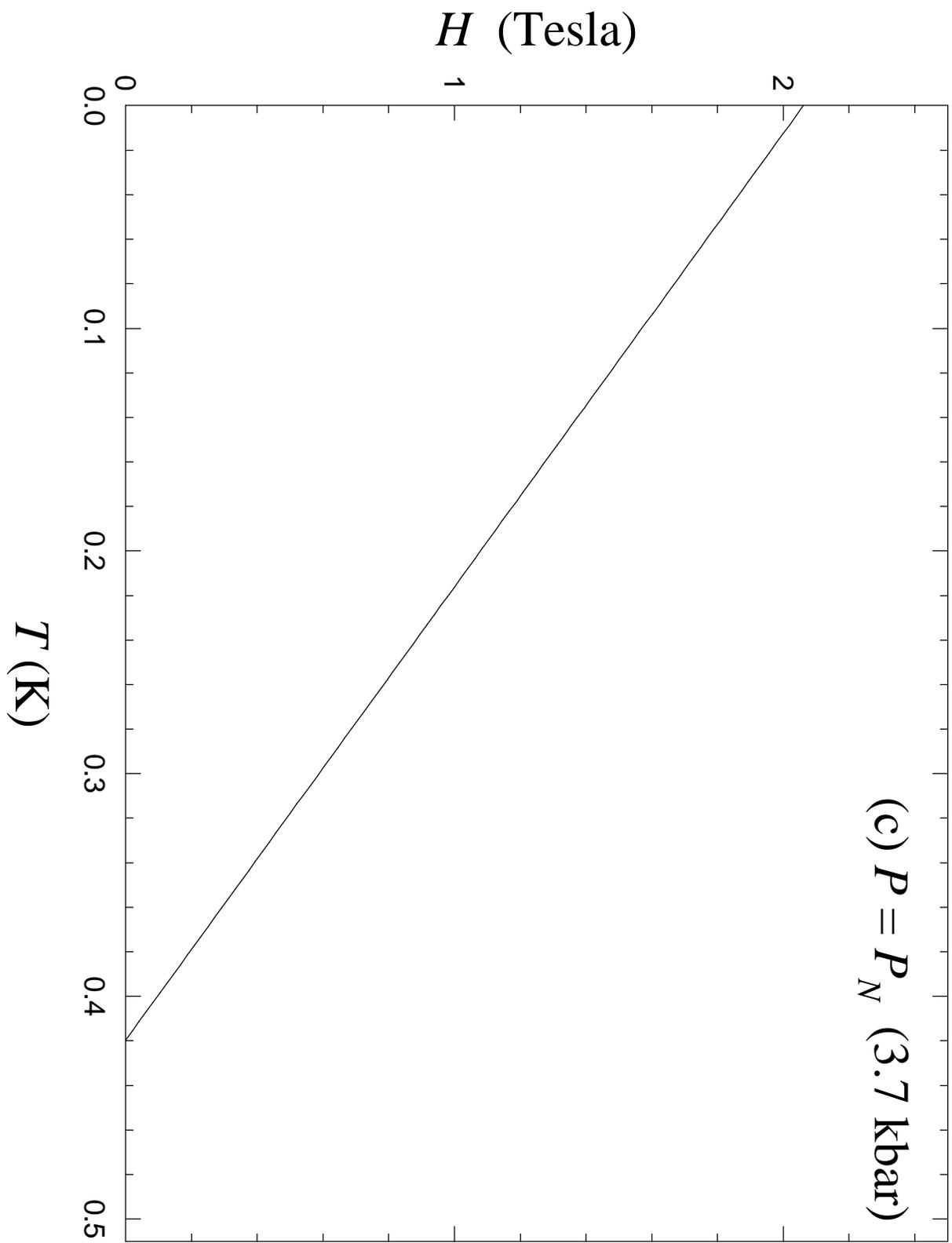
\author{K. A. Park and Robert Joynt}
\address{
Department of Physics and Applied Superconductivity Center\\
University of Wisconsin-Madison\\
1150 University Avenue\\
Madison, Wisconsin 53706\\}
\date{\today}
\maketitle
\begin{abstract}
The phase diagram of superconducting UPt$_3$ is explained in a
Ginzburg-Landau theory starting from the
hypothesis that the order parameter is a pseudo-spin
singlet which transforms according to
the $E_{1g}$ representation of the $D_{6h}$ point group.
We show how to compute the positions of the phase boundaries
both when the applied field is in the basal plane and when
it is along the c-axis.  The experimental phase diagrams as
determined by longitudinal sound velocity data can be fit
using a single set of parameters.
In particular the crossing of the upper critical field curves
for the two field directions and the apparent isotropy of the
phase diagram are
reproduced.  The former is a result of the magnetic
properties of UPt$_3$ and their contribution to the
free energy in the superconducting state. The latter is a consequence
of an approximate particle-hole symmetry.  Finally we extend the
theory to finite pressure and show that, in contrast to
other models, the $E_{1g}$ model explains the observed
pressure dependence of the phase boundaries.

\end{abstract}
\pacs{PACS Nos. 74.70.Tx, 74.25.Dw, 74.20.De.}
\narrowtext
\section{Introduction}
\label{intro}
Currently, there is a great deal of discussion about the nature of the
superconducting heavy-fermion compunds, especially UPt$_3$.  Much of this
discussion has centered on the unusual nature of the superconducting state.
Experiments to map out the phase diagram of UPt$_3$ in the
field-temperature ($H - T$) plane using both specific heat
\cite{fisher,hasselbach} and longitudinal
sound absorption \cite{berlin,mil} and velocity \cite{adenwalla,lin}
have revealed multiple superconducting
phases.  In particular
these measurements show that two superconducting phases exist even at zero
field, as was predicted \cite{joynt88} by an analysis of the free energy for
a two-component order parameter in the presence of antiferromagnetism.
\cite{aeppli}
The resulting Ginzburg-Landau (G-L) theory makes additional
predictions - e.\,g.\, the kink in the upper critical field when the
field is in the basal plane.\ \cite{hess,mach89} In these theories,
the order parameter transforms as one of the
irreducible representations of the $D_{6h}$ point group of the crystal.
either $E_1$ or $E_2$.\cite{volovik,sigrist}

Further evidence about the superconducting state of UPt$_3$ comes from
measurements of ultrasound \cite{shivaram,hirschfeld} and heat
conduction. \cite{lussier}  These experiments suggest
that there are point nodes in the superconducting gap function where the
Fermi surface intersects the line $k_x = k_y = 0$ and line nodes where the
Fermi surface intersects the $k_z = 0$ or $k_z = \pi/c$ planes.  This is
evidence for a
d-wave $E_{1g}$ order parameter which transforms like $(k_x k_z, k_y k_z)$.
The theorem of Blount \cite{Blount1} states that triplet states cannot have
lines of nodes when spin-orbit coupling is taken into account.  The theorem
assumes that no
symmetries are present other than the crystal point group symmetries.
It has been argued that other symmetries may be present in UPt$_3$
\cite{adphys}
and thus lines of nodes may be present even if the Cooper pair is a triplet.
Thus the nodal pattern may not prove singlet pairing.

In spite of the success of the $E_{1g}$ model in explaining the nodal
structure of the gap function and the existence of multiple superconducting
phases, certain objections have been raised regarding its suitability as a
description of UPt$_3$.  One objection is that the $E_{1g}$ theory fails to
explain the isotropy of the phase diagram, or, in other words, why the
phase diagram when the field is parallel to the c-axis of the crystal
appears to be similar to the phase diagram when the field is perpendicular
to the c-axis. \cite{adphys,garg}  We shall argue that there are similarities
but also important differences and
that the $E_{1g}$ theory does in fact explain the phase diagram for both
orientations of $\bbox{H}$.  The other common objection to the $E_{1g}$
theory is that because it is a singlet theory it cannot explain why the upper
critical field curve for $\bbox{H}$ along the c-axis
and the curve for $\bbox{H}$ in the basal plane cross.
\cite{shivaram2}  This
crossing is maintained to be a characteristic of triplet theories alone.
\cite{choi} By
a careful analysis of the magnetic properties of UPt$_3$ and their
contributions to
the G-L free energy we will show that pseudo-spin singlet
states can also produce this effect.

The plan for the rest of this paper is as follows.  In Sec.\ \ref{math} the
overall mathematical approach to the phase diagram problem is discussed.
It is necessary to go into the method in some detail: only a very careful
analysis brings out the nature of the inner phase transition.  In
Sec.\ \ref{basal} we will take the free energy and use it to obtain the phase
diagram when $\bbox{H}$ is
in the basal plane.  The observed tetracritical point comes out
in a natural way.  By fitting the theory to the longitudinal velocity
data we will obtain values for all the relevant parameters of our theory.
Then in Sec.\ \ref{caxis} we will obtain the phase diagram for the case
when $\bbox{H}$ is parallel to the c-axis.  We will show that our theory
can be fit to the data for the case when $\bbox{H}$ is parallel to the
c-axis with the same set of parameters used for the case when $\bbox{H}$ is
in the basal plane.  The near-crossing of the phase boundaries when
$\bbox{H}$ is along the c-axis
is a consequence of approximate particle-hole symmetry. In Sec.\
\ref{magnetic} we will discuss the magnetic properties, in particular the
magnetic susceptibility.  We will show the effect the susceptibility of UPt$_3$
has on the G-L free energy and how this leads to properties such as the
crossing of the upper critical field curves for different directions of the
field.   In Sec.\ \ref{press} the phase diagram is extended to finite pressure.
Finally in Sec.\ \ref{ending} we make some concluding remarks.

\section{Effective field method for the phase diagram}
\label{math}
This section will be devoted to explaining the mathematical
method used to obtain the phase diagram for UPt$_3$ in the presence of
an external magnetic field.  The full problem is very
complicated.  We give first a simple example to orient the
reader to the case of competing order parameters.  The
reader who is mainly interested in the overall concept,
not the details, may read the first subsection and consult the
summary figures in the other subsections.

\subsection{Simple model}
A simple system with a multicomponent order parameter
and competition among the components is a magnet with uniaxial anisotropy.
The free energy is:

\begin{eqnarray}
F&=&\alpha_{0x} (T-T_c) (M_x^2 + M_y^2) + \beta_{xy}(M_x^2 + M_y^2)^2 \\
 & &\mbox{} + \alpha_{0z} (T-T_z) M_z^2 + \beta_z M_z^4 + \beta_{xz}
(M_x^2 + M_y^2) M_z^2.  \nonumber
\end{eqnarray}
Suppose that $T_c > T_z$.  Then at $T_c$, the system develops
a nonzero $M$ in the $x$-$y$ plane, its direction otherwise not determined by
$F$. Let us say $\bbox{M} = M\bbox{\hat{x}}$ with $M(T)$ given by
$<M^2>= \alpha_{x0} (T_c-T)/ 2 \beta_{xy}$.  The angle brackets indicate
equilibrium values.  The question we face (which adumbrates the whole theme
of this paper) is: how do $M_y$ and $M_z$ behave below $T_c$ ?  The first
question has a simple
answer.  $M_y$ will remain
zero below $T_c$.  One way to see this is to write an effective free energy
for $M_y$ below $T_c$ by simply taking the terms in $F$ which involve $M_y$
and writing the equilibrium value for $M_x$ and $M_z$:
\begin{eqnarray}
F_{\text{\em eff}}(M_y) & = & \alpha_{0x} (T-T_c) M_y^2 + 2 \beta_{xy}
<M_x^2> M_y^2 + \beta_{xz} <M_z^2> M_y^2 + \beta_{xy} M_y^4 \\
& = &  \beta_{xz} <M_z^2> M_y^2 + \beta_{xy} M_y^4.
\end{eqnarray}
There will be a temperature range below $T_c$ where $<M_z^2> = 0$.
The fact that the effective free energy is then quartic in $M_y$
is the sign that $M_x$ and $M_y$ are degenerate, and the fact
that the minimum of $F_{\text{\em eff}}$ is at $M_y^2 = 0$ indicates that
rotation of $\bbox{M}$ in the $x$-$y$ plane
will take place only if a magnetic field (which could be infinitesimal)
is applied.

Now do the same for $M_z$:
\begin{eqnarray}
F_{\text{\em eff}}(M_z) &=& \alpha_{0z} (T-T_z) M_z^2 + \beta_{xz} <M_x^2>
                               M_z^2  + \beta_{z} M_z^4 \\
            &=& [ \alpha_{0z} (T-T_z) + \frac{\beta_{xz}}{2 \beta_{xy}}
                \alpha_{0x} (T-T_c) ] M_z^2  + \beta_{z} M_z^4
\end{eqnarray}
There are evidently two possibilities.  Either the expression in square
brackets
vanishes at positive $T$, in which case there is a second-order phase
transition where $M_z$ appears so that the magnetization rotates in the
$x$-$z$ plane, or it vanishes at negative $T$, which implies
that there is no further transition and $M_z = 0$ at all $T$.
The rotational phase transition, which is second-order,
takes place at a lower critical
temperature given by
\begin{equation}
T_{c2}=\frac{\alpha_{0z}T_z+(\beta_{xz}\alpha_{0x}/2 \beta_{xy})T_c}
            {\alpha_{0z}+(\beta_{xz}\alpha_{0x}/2 \beta_{xy})}.
\label{eq:magtc}
\end{equation}
An important point is that $T_{z}$, the {\it bare} critical temperature
for $M_z$ may be positive but $T_{c2}$ still negative.  This would be an
example of the effective field suppressing a transition.
If there is a transition ($T_{c2}>0$), then the effective free
energy is not valid for $T<T_{c2}$ - it neglects the
feedback of $M_z$ on $M_x$.

Of interest below will be the question of artificial
terms such as $ \gamma M_x^3 M_z$ in the
original free energy.  This would add a term
$ \gamma [\alpha_{0x} (T_c - T) / 2 \beta_x ]^{3/2} M_z$ to
the effective free energy for $M_z$.  This means that $M_z$
becomes nonzero already at $T_c$ and the lower transition is
converted to
a crossover, just as if an external field in the $z$-direction
were applied.

For finding out whether there is a
transition, whether it is second order, and computing the
lower transition temperature, analysis of the effective free energy
is all that is required.  To find the behavior of the system below
$T_{c2}$, one must minimize of the full free energy.

\subsection{s-wave superconductor}

We now apply the effective field method
to the well-known problem of an
isotropic s-wave superconductor to show how it
works in a case which is actually quite non-trivial, but whose
phase diagram is well understood.
This system has a single
complex order parameter.  In the presence of a field, however,
there are, in a certain sense, an infinite number of
order parameters, and interesting competition among them.

The free energy density for the system is
\begin{equation}
f=\alpha_0(T-T_c) \mid \eta \mid ^2
  + \beta \mid \eta \mid ^4
  + K \sum_i D_i \eta D_i^* \eta^*.
\end{equation}
Here $D_i = -i  \partial_i + 2eA_i / \hbar c$ ($-e$ is the charge on an
electron) and
if we take take $\bbox{H}$ in the $z$-direction, then
the gauge $\bbox{A}=Hx \bbox{\hat{y}}$ is appropriate.
We have $D_x = -i  \partial_x$ and
$D_y = -i  \partial_y + 2eHx / \hbar c$.
Our problem is to minimize the free
energy $F=\int f dV$ for
arbitrary $H$ and $T$.

The method we will use is to expand
the function $ \eta(\bbox{x}) $ in a complete set
of normalized eigenfunctions of the operator
\begin{equation}
K (D_x^2+D_y^2),
\end{equation}
which are
\begin{equation}
\phi_{nk}=[\ell/\pi L_y^2]^{1/4} e^{-iky}\exp[-(x-k\ell^2)^2/2\ell^2]
H_n((x-k\ell^2)/\ell),
\end{equation}
where $\ell = \hbar c / 2 e H$, $H_n$ are the Hermite polynomials,
and $L_y$ is the size of the system in the
$y$-direction.
We now write
\begin{equation}
\eta(\bbox{x})=\sum_{nk} C_{nk} \phi_{nk} (\bbox{x})
\end{equation}
and the free energy
becomes
\begin{equation}
F=\sum_{nk} [\alpha_0 (T-T_c) + \varepsilon_{n}] \mid C_{nk} \mid^2
+ \sum_{n_{1}k_1,n_2k_2,n_3k_3,n_4k_4}
   b_{n_{1}k_1,n_2k_2,n_3k_3,n_4k_4}
    C_{n_1k_1}C_{n_2k_2}^*C_{n_3k_3}C_{n_4k_4}^*.
\label{eq:poly}
\end{equation}
The coefficients in this equation are:
\begin{equation}
\varepsilon_n = (n+1/2) \frac{4 K e H}{\hbar c}
\end{equation}
and
\begin{equation}
b_{n_{1}k_1,n_2k_2,n_3k_3,n_4,k_4}
= \beta \int d^3x
    \phi_{n_1k_1}\phi_{n_2k_2}^*\phi_{n_3k_3}\phi_{n_4k_4}^*.
\end{equation}
An important point is that $\varepsilon$ is
independent of $k$ and
$b$ is zero unless $k_1-k_2+k_3-k_4 =0$.
We have re-expressed $F$ as a fourth-order polynomial
in an infinite number of variables $C_{nk}$, which may
be thought of formally as competing order parameters.
We must minimize this polynomial.

The upper critical field curve is given by noticing when
the coefficient of the quadratic term {\it first} changes sign:
\begin{equation}
\alpha_0 (T-T_c) + \varepsilon_{n} = 0;
\label{eq:hc2}
\end{equation}
The highest value of $H$ for which this equation holds
corresponds to $n = 0$ and the curve
\begin{equation}
\alpha_0 (T-T_c) + 2 K e H / \hbar c = 0
\end{equation}
defines the normal-- superconducting phase boundary.

Below this boundary, some {\it but not all}
of the $C_{nk}$ are nonzero and
\begin{equation}
C_{0k} \sim [\alpha_0 (T_c-T) - 2 K e H / \hbar c]^{1/2} = \delta^{1/2}.
\end{equation}
This equation defines $\delta$, which serves as a small
quantity in the analysis below, the validity of which is thereby
limited to the neighborhood of the phase boundary.
$\delta > 0$ in the ordered phase.
The periodicity of the
flux lattice shows that $C_{0k} \neq 0 $
if and only if $k = mq$, where $m$ is any integer
and $q = \sqrt{\sqrt{3} \pi} / \ell$.  We shall denote
this condition by $ k \in L$, i.\,e.\, $k$ belongs to
the discrete set which constitutes the flux lattice.
The discreteness reflects the fact that
magnetic translation symmetry as well as gauge
symmetry are broken in the low-$T$, low-$H$ phase.
Thus, sufficiently close to the phase boundary,
only these coefficients need be computed and we get
the familiar theory of the hexagonal flux lattice.
As is well known, no further phase transitions take
place as the field is lowered until the Meissner
state takes over at $H_{c1}$.

In UPt$_3$, on the other hand, there is another transition
when the field is reduced.
Why does this not occur in the s-wave case?
The answer is not obvious.  For example, we may consider the Landau level $n =
1$.
Setting the eigenvalue equal to zero as we did for
$n = 0$ would give a critical field line with the
same $T_c$ but with a slope only 1/3 of the slope of
the $H_{c2}$ curve.  Why does no transition take place on this line
in the $H-T$ plane?
That is, why is there no nonalytic behavior of the
$C_{1k}$ on this line?  What about $C_{0k}$ for $k \neq mq$ ?

To answer these questions, we must develop a picture of the
effective fields present in the system when the symmetry
has been broken.  This is done by
classifying different terms of the polynomial
in Eq.\ \ref{eq:poly}.

\subsubsection*{Class 1: terms determing the leading
behavior of $C_{0k}$, $k \in L$.}
These are the simplest of all;
the free energy is
\begin{equation}
F=\sum_{k} [\alpha_0 (T-T_c) + \varepsilon_{0}] \mid C_{0k} \mid^2
+ {\cal O} (C_{0k}^4) + \ldots,
\label{eq:p0k}
\end{equation}
where only the terms relevant to
the behavior of terms in class 1 have been written explicitly.

For small $\delta$, these terms give the simple result
\begin{equation}
F = - \delta (C1)^2 + {\cal O} ((C1)^4) \Rightarrow (C1) \sim \delta^{1/2},
\end{equation}
where $(C1)$ denote collectively the $C_{nk}$ which belong to
class 1.  For our considerations which are simply a
matter of power counting, the indices on $C$ are not required at this point.
The $C1$ are analogous to $M_x$ in the magnetic example.
The conclusion is that the $C_{nk}$ are proportional
to $\delta^{1/2}$ near the phase boundary.

\subsubsection*{Class 2: terms determining the leading behavior of
$C_{0k'}$, $k' \in L' $.}
  We write momenta
of the form $ k' = ( m+ \frac{1}{2}) q $, where
$m$ is an integer,
with a prime.  Combining the $C_{0k'}$ builds a
hexagonal lattice which interpenetrates the original one,
as we shall see below in section \ref{basal}.  When $\delta = 0$
these variables are degenerate with the $(C1)$ -
these are the ones not chosen because of the
breaking of the magnetic translation symmetry.
The $C2$ should be compared to $M_y$ in the previous subsection.
As $\delta$ increases, they become less favored
because they feel an effective repulsion from the $(C1)$.
The relevant terms in $F$ (call them collectively $F_{C1,C2}$)
are of the form:
\begin{eqnarray}
F_{C1,C2} & = & \sum_{k'} [\alpha_0 (T-T_c) + \varepsilon_{0}] \mid C_{0k'}
\mid^2
\label{eq:p0kp}   \\
  &   & \mbox{} + \sum_{k_1,k_2,k_3,k_4}
   b_{0k_1,0k_2,0k_3,0k_4}
    C_{0k_1}C_{0k_2}^*C_{0k_3}C_{0k_4}^* + \ldots. \nonumber
\end{eqnarray}
This equation
repays careful examination.
A first crucial point is that
there are no terms of the form
$(C1)^3 (C2)$ or $(C1) (C2)^3$.  Recall that
$b_{n_{1}k_1,n_2k_2,n_3k_3,n_4k_4}$ is zero
unless $k_1 - k_2 + k_3 - k_4 = 0$.
However the $k \in L$ are equally spaced,
so if $k_1,k_2,k_3 \in L$, then $k_4 \in L$ as well.
Similarly if $k_1,k_2,k_3 \in L'$, then also $k_4 \in L'.$
For the case $k_1, k_2 \in L$ then we can have $k_3, k_4 \in L'$.
Hence the only cross terms (in $L$ and $L'$) which survive have the form
$(C1)^2 (C2)^2$.
More explicitly,
\begin{eqnarray}
F_{C1,C2} & = & \sum_{k'} [\alpha_0 (T-T_c) + \varepsilon_{0}] \mid C_{0k'}
\mid^2
+  \sum_{k_1,k_2,k_3',k_4'} B_{0k_1,0k_2,0k_3',0k_4'}
C_{0k_1}C_{0k_2}^*C_{0k_3'}C_{0k_4'}^*   \\
&  & \mbox{} + \sum_{k_1,k_2,k_3',k_4'} b_{0k_1,0k_3',0k_2,0k_4'}
C_{0k_1}C_{0k_2}C_{0k_3'}^*C_{0k_4'}^* + c.c. \nonumber   \\
&  & \mbox{} + \sum_{k_1',k_2',k_3',k_4'}
   b_{0k_1',0k_2',0k_3',0k_4'}
    C_{0k_1'}C_{0k_2'}^*C_{0k_3'}C_{0k_4'}^* + \ldots. \nonumber
\end{eqnarray}
In this equation
\begin{equation}
B_{0k_1,0k_2,0k_3',0k_4'}=b_{0k_1,0k_2,0k_3',0k_4'}+b_{0k_1,0k_4,0k_3',0k_2'} +
                        b_{0k_3,0k_4,0k_1',0k_2'} + b_{0k_3,0k_2,0k_1',0k_4'}.
\end{equation}
In the summations $k$ runs over $L$ and $k'$ runs over $L'$.
The idea of the effective field is to note that,
when $H < H_{c2}$ (or $\delta >0$),
we may write an effective free energy for the $(C2)$:
\begin{eqnarray}
F_{{\text{\em eff}}}((C2)) & = & \sum_{k'} [\alpha_0 (T-T_c) +
\varepsilon_{0}] \mid C_{0k'} \mid^2 \\
& & \mbox{} +   \sum_{k_1,k_2,k_3',k_4'} B_{0k_1,0k_2,0k_3',0k_4'}
<C_{0k_1}C_{0k_2}^*>C_{0k_3'}C_{0k_4'}^*   \nonumber \\
&  & \mbox{} + \sum_{k_1,k_2,k_3',k_4'} b_{0k_1,0k_3',0k_2,0k_4'}
<C_{0k_1}C_{0k_2}>C_{0k_3'}^*C_{0k_4'}^* + c.c. \nonumber  \\
& & \mbox{} + \sum_{k_1',k_2',k_3',k_4'}
   b_{0k_1',0k_2',0k_3',0k_4'}
    C_{0k_1'}C_{0k_2'}^*C_{0k_3'}C_{0k_4'}^* + \ldots, \nonumber
\end{eqnarray}
where the angle brackets denote equilibrium values in the
ordered phase.  Examination of this free energy
is all that is required to analyze the stability of $L'$.  Since $C_{0k} \sim
\delta^{1/2}$ the structure of this equation is
\begin{equation}
F_{{\text{\em eff}}}((C2))= - \delta(1+R2) (C2)^2 + {\cal O} ((C2)^4),
\end{equation}
where $R2$ is a dimensionless matrix which is independent of
temperature.  In fact if the $C_{0k}$ are divided into real and
imaginary parts then $R2$ is a real, symmetric matrix.
{\it If there is to be no further
phase transition, then all the eigenvalues of $R2$ must be
less than or equal to -1.}  Otherwise the $(C2)$ condense, and another
lattice would form.  This would mean an ``inner'' transition
in ordinary s-wave materials, which does not occur.
Of course the $L'$ lattice is degenerate with the $L$ lattice.
One may be converted into the other
with application of an infinitesimal external current
as $M_x$ may be converted to $M_y$ in the simple model
by an infinitesimal external magnetic field.

\subsubsection*{Class 3: terms determining the behavior of
$C_{nk}$ for $n > 0 $ and $k \in L.$}
Two subcases must be distinguished here: $n$ = even and $n$ = odd.
If $n$ is even then all possible terms come into
the effective free energy for the $(C3)$:
\begin{eqnarray}
F_{{\text{\em eff}}}((C3)) & = & [\alpha_0 (T-T_c) + (2n+1) 2 K e H /
\hbar c] (C3)^2 + b3_1 <(C1)^3> (C3)  \\
& & \mbox{} + b3_2 <(C1)^2> (C3)^2 + b3_3 <(C1)^3> (C2)  \nonumber
\end{eqnarray}
where $F_{{\text{\em eff}}}$ has been written for a definite $n$ value.
The coefficient in square brackets is positive as long as
$H > H_{c2}/ (2n+1) $.  $ b3_1, b3_2, $ and $b3_3$ are constants
independent of $\delta$ whose precise
form need not detain us.  In this field region the leading
behavior is dominated by the term linear in the
$(C3)$:
\begin{equation}
F_{{\text{\em eff}}}((C3)) \sim   - (C3)^2 + b3_1 (C1)^3 (C3)
 \sim  -(C3)^2 + b3_1 \delta^{3/2} (C3) \Rightarrow (C3) \sim \delta^{3/2}.
\end{equation}
This resolves the question raised above.  There
is no phase transition at $H_{c2}/(2n+1)$ for $n$ even
because the
coefficients determining the weight of the $n$th
Landau level have already started to grow at
$H_{c2}$ itself.  Thus the putative phase transition is
converted into a crossover.

For $n$ = odd,
only even terms in $(C3)$ appear for parity reasons.
These remain zero below $H_{c2}$.  One argues in the
same way as in case 2; since the $(C3)$ are less stable
than the $(C2)$ because of the higher bare quadratic coefficient
in front, the conclusion follows {\it a fortiori}.

\subsubsection*{Class 4: terms determining the leading behavior of
$C_{nk'}$, $n > 0$, $k' \in L' $.}
We have accumulated enough experience to write the
effective free energy immediately:
\begin{eqnarray}
F_{{\text{\em eff}}}((C4)) & = & \sum_{nk'} [\alpha_0 (T-T_c) +
(2n+1) 2 K e H / \hbar c] \mid C_{nk'} \mid^2  \\
& & \mbox{} +  16 \sum_{k_1,k_2,nk_3',n'k_4'} b_{0k_1,0k_2,nk_3',n'k_4'}
<C_{0k_1}C_{0k_2}^*>C_{nk_3'}C_{n'k_4'}^*   \nonumber \\
&  & \mbox{} + 16 \sum_{k_1,k_2,nk_3',n'k_4'} b_{0k_1,0k_3',nk_2,n'k_4'}
<C_{0k_1}C_{0k_2}>C_{nk_3'}^*C_{n'k_4'}^* + c.c. \nonumber \\
& & \mbox{} + \sum_{k_1',k_2',nk_3',n'k_4'}
   b_{0k_1',0k_2',nk_3',n'k_4'}
    C_{0k_1'}C_{0k_2'}^*C_{nk_3'}C_{n'k_4'}^* + \ldots. \nonumber
\end{eqnarray}
By analogy with case 2, this may be written as
schematically as
\begin{equation}
F_{  \text{\em eff}  }((C4))= - \delta(1+R4) (C4)^2 +{\cal O}( (C4)^4),
\end{equation}
the only difference being that $R4$ is a matrix in the
$n$ index as well as the $k'$.  There are no terms linear
in the $(C4)$.  There is no phase transition
involving the $(C4)$ - hence the eigenvalues of $R4$ are less than -1.
The $(C4)$ are always zero in equilibrium.

\subsubsection*{Class 5: terms determing the behavior of
other $k$ values.}
  It is evident from the momentum
conservation condition that the effective free energy
for $k$ such that $k \not\in L$, $k \not\in L' $
can contain terms such as $ <(C1)> (C5)^3 $, for
$k = q/3$, for example.  There are no terms linear in the $C5$.
The cubic terms could give rise
to additional first order transitions in principle.
It is evident that this does not occur and we do not consider such terms
further.

We may conclude this discussion of the s-wave case by a graphical account of
how
all transitions except the one at $H_{c2}$ itself are suppressed.
Fig.\ \ref{fig:htswave} shows the bare eigenvalue curves,
the repulsion of some levels away from the $H_{c2}$
curve and the conversion to crossover of others.
These two effects arise from the effective field coming from the quartic term
in the
original free energy.

\subsection{Effective field method for the d-wave case}

The d-wave case may be analyzed in a similar manner.
For our present qualitative discussion we need only the
fact that the order parameter becomes a complex
two-component vector $\bbox{\eta} = (\eta_x, \eta_y)$.
The free energy again contains quadratic and quartic terms in this
variable.  It is:
\begin{eqnarray}
\label{eq:df}
f &=& \alpha_0 (T-T_x) |\eta_x|^2 + \alpha_0 (T-T_y) |\eta_y|^2 +
    \beta_1 ( \bbox{\eta} \cdot \bbox{\eta}^*)^2 +
    \beta_2 | \bbox{\eta} \cdot \bbox{\eta}|^2  \\
  & & \mbox{} + \sum_{i,j=x,y} ( K_1 D_i \eta_j D_i^* \eta_j^* +
  K_2 D_i \eta_i D_j^* \eta_j^* + K_3 D_i \eta_j D_j^* \eta_i^*) +
  K_4 \sum_{i=x,y} |D_z \eta_i|^2 \nonumber
\end{eqnarray}
We have neglected certain terms which are not relevant to the
present discussion.  They will be introduced in the next section.
However, we do not specialize to any particular direction of
field, and the analysis is valid for all directions.
This section generalizes the analysis carried out
by Joynt \cite{joynt91} for the field in the basal plane,
which is the easiest case.

The quadratic form may be diagonalized by finding the
two-component vector eigenfunctions $\bbox{\phi_{nk}}$. We then write
\begin{equation}
\bbox{\eta} = \sum_{nk} C_{nk} \bbox{\phi_{nk}}
\end{equation}
to obtain
\begin{eqnarray}
F & = & \sum_{nk} [\alpha_0(T-T_{c1})+\varepsilon_{n}(H)] \mid C_{nk} \mid^2
\label{eq:dpoly} \\
& & \mbox{} + \sum_{n_{1}k_1,n_2k_2,n_3k_3,n_4k_4}
   b_{n_{1}k_1,n_2k_2,n_3k_3,n_4k_4}
    C_{n_1k_1}C_{n_2k_2}^*C_{n_3k_3}C_{n_4k_4}^*, \nonumber
\end{eqnarray}
the difference with the s-wave case being that
the energy levels  $\varepsilon_{n}(H)$
and the form of the $b$ coefficients are
far more complicated.  Here $T_{c1}$ is the greater of $T_x$ and $T_y$, and
$T_{c2}^0$ is the lesser of $T_x$ and $T_y$.
The curves which are the lines  $\alpha_0(T-T_{c1})+\varepsilon_{n}(H) = 0$
are shown in Fig.\ \ref{fig:htdwave}(a).
Crucially, however, the momentum conservation
condition
$k_1-k_2+k_3-k_4 =0$
is the same.  We now specialize a bit to the case of UPt$_3$.
Then the solutions $\alpha_0(T-T_{c1})+\varepsilon_{n}(H)=0$
fall into two classes.  Half the levels
have a {\it bare} $T_c$ at $T_{c1}$ [because $\varepsilon_{n}(H=0) = 0$] and
half at $T_{c2}^0$ [because $\varepsilon_{n}(H=0) =
\alpha_0(T_{c1}-T_{c2}^0)$].
Thus when we specify the level index it must be stated to which class the
level belongs.  Call those with the higher $T_c$ (a) and those with the
lower $T_c$ (b).
Apart from this difference, the classification of states proceeds similarly to
the s-wave case.

\subsubsection*{ 1(a): terms determining the leading behavior of
$C_{n_ak}$, $n = 0$, $k \in L $.}
The free energy is
\begin{equation}
F(C_{0_ak}) = \sum_{k}[\alpha_0(T-T_{c1})+ \varepsilon_{0_a}] \mid C_{0_ak}
 \mid^2 + {\cal O} (C_{0_ak}^4) + \ldots,
\label{eq:p0kd}
\end{equation}
with the result that
\begin{equation}
- \delta (C1)_a^2 + {\cal O} ((C1)_a^4) \rightarrow (C1)_a \sim \delta^{1/2},
\end{equation}
where $\delta = -[\alpha_0(T-T_{c1})+\varepsilon_{0_a}]$.

\subsubsection*{Class 2(a): terms determining the leading behavior of
$C_{0_ak'}$, $k' \in L' $.}
  Again, we write momenta
of the form
$( m+ \frac{1}{2}) q $, where
$m$ is an integer,
with a prime.  These coefficients may be treated by
analogy with the s-wave class 2 above.  Familiar with the
procedure, we may write down the relevant effective free energy
immediately:
\begin{equation}
F_{{\text{\em eff}}}((C2)_a)= - \delta(1+R2_a) (C2)_a^2 + {\cal O}
((C2)_a^4),
\end{equation}
where again $R2_a$ is a dimensionless matrix.  This matrix is similar
to $R2$, and we expect that all the eigenvalues of $R2_a$
must be
less than or equal to -1 and the usual hexagonal symmetry arises at $H_{c2}$.
That this is actually the case has been shown by Luk'yanchuk and Zhitomirskii.
\cite{zhito}.
\subsubsection*{Class 3(a): terms determining the behavior of
$C_{n_ak}$ for $n > 0 $ and $k \in L.$}
The analysis proceeds as in s-wave class 3.
\begin{equation}
F_{{\text{\em eff}}}((C3)_a) \sim  (C3)_a^2 + b3_{1a} (C1)_a^3 (C3)_a
\sim (C3)_a^2 + b3_{1a} \delta^{3/2} (C3)_a \rightarrow (C3)_a \sim
\delta^{3/2}.
\end{equation}
These candidate phase transitions are thus converted to crossovers by the
effective field.  Note that there is, for general field directions, no parity
selection
rule in the d-wave case so there is no distinction between $n$ = odd and
$n$ = even.
The $(C3)_a$ contribute to the change of
shape of the vortices as as the external field is reduced
below $H_{c2}$, but produce no further phase transition.

\subsubsection*{Class 4(a): terms determining the leading behavior of
$C_{n_ak'}$, $n_a > 0$, $k' \in L' $.}
The effective free energy is:
\begin{equation}
F_{{\text{\em eff}}}((C4)_a)= - \delta(1+R4_a) (C4)_a^2 + {\cal O}
((C4)_a^4).
\end{equation}
We again argue by analogy with s-wave case 4
that the eigenvalues of $R4_a$ must be less than -1 so that
the $(C4)_a$ are always zero.

\subsubsection*{Class 5(a): other periodicities.}
Again other periodicities will not arise from states in
class a, just as in s-wave, case 5.

In discussing cases 1(a) to 5(a), we stress that
we have not given explicit proofs for the magnitude
of the eigenvalues of the different $R$-matrices.
It is possible to write these matrices formally,
but they are quite complicated.   However, they are very
similar to the s-wave case, where we
are certain of the result even in the absence of
explicit calculation.

Let us now turn to the the levels which start from $T_{c2}^0$.
Here we have less guidance
from the s-wave analogy.

\subsubsection*{Class 1(b): terms determing the leading
behavior of $C_{0_bk}$, $k \in L$.}
These terms are analogous to those in
s-wave class 3.
As long as $H \neq 0 $, all possible terms come into
the effective free energy for the $(C1)_b$:
\begin{eqnarray}
F_{{\text{\em eff}}}((C1)_b) & = & [\alpha_0(T-T_{c1}) +
\varepsilon_{0_b}] (C1)_b^2 +
b3_{1ab} <(C1)_a^3> (C1)_b \\ & & \mbox{} + b3_{2ab} <(C1)_a^2> (C1)_b^2
 + b3_{3ab} <(C1)_a^3> (C1)_b. \nonumber
\end{eqnarray}
For certain special directions of the field,
the cubic-linear and linear-cubic terms may vanish,
but we are concerned here with the general case.
This leads to the result that
\begin{equation}
F_{{\text{\em eff}}}((C1)_b) \sim  (C1)_b^2 + b3_{1ab} (C1)_a^3 (C2)_b
\sim (C1)_b^2 + b3_{1ab} \delta^{3/2} (C1)_b \rightarrow (C1)_b \sim
\delta^{3/2}.
\end{equation}
Thus these terms show crossover behavior.
This fact is the apparent basis for a statement occasionally found
in the literature that for general directions of the field there
is no lower phase transition in d-wave systems at
finite field \cite{varma,garg}.

\subsubsection*{Class 2(b): terms determining the leading behavior of
$C_{0k'}$, $k' \in L' $.}
  This is the
crucial case so we treat it in detail.
The relevant terms in $F_{{\text{\em eff}}}$ are:
\begin{eqnarray}
F_{{\text{\em eff}}}((C2)_{b}) & = & \sum_{k'} [\alpha_0(T-T_{c1}) +
\varepsilon_{0_b} ] \mid C_{0_bk'} \mid^2 \\
& & \mbox{} + \sum_{k_1,k_2,k_3',k_4'} B_{0_ak_1,0_ak_2,0_bk_3',0_bk_4'}
<C_{0_ak_1}C_{0_ak_2}^*>C_{0_bk_3'}C_{0_bk_4'}^*  \nonumber \\
&  & \mbox{} + \sum_{k_1,k_2,k_3',k_4'} b_{0_ak_1,0_bk_3',0_ak_2,0_bk_4'}
<C_{0_ak_1}C_{0_ak_2}>C_{0_bk_3'}^*C_{0_bk_4'}^* + c.c. \nonumber \\
& & \mbox{} + \sum_{k_1',k_2',k_3',k_4'}
   b_{0_ak_1',0_ak_2',0_bk_3',0_bk_4'}
    <C_{0_ak_1'}C_{0_ak_2'}^*>C_{0_bk_3'}C_{0_bk_4'}^* + \ldots,  \nonumber
\end{eqnarray}
where the angle brackets denote equilibrium values in the
ordered phase and
\begin{eqnarray}
B_{0_ak_1,0_ak_2,0_bk_3',0_bk_4'} &=& b_{0_ak_1,0_ak_2,0_bk_3',0_bk_4'} +
                                    b_{0_ak_1,0_bk_4,0_bk_3',0_ak_2'} \\
                                  & &    +b_{0_bk_3,0_bk_4,0_ak_1',0_ak_2'} +
                                    b_{0_bk_3,0_ak_2,0_ak_1',0_bk_4'} \nonumber
\end{eqnarray}
Since $C_{0_ak} \sim \delta^{1/2}$
the structure of this equation is
\begin{equation}
F_{{\text{\em eff}}}((C2)_b)= [\alpha_0(T-T_{c1}) + \varepsilon_{0_b}
- \delta R2_b]
(C2)_b^2 + {\cal O} ((C2)_b^4),
\end{equation}
where again $R2_b$ is a dimensionless matrix.
The question of further phase transitions in this case
boils down to asking whether the matrix
$ [\alpha_0(T-T_{c1}) + \varepsilon_{0_b}- \delta R2_b]$ can ever have
negative eigenvalues.  If it does, then there will be
a lower phase transition.  The answer is known for
UPt$_3$ (more formally for $E$ representations of
the hexagonal group in some parameter ranges) in certain limiting cases.  If
$H =0$, then
the problem reduces to a well-known one
\cite{sjr,hess,mach89}.  There are indeed two
transitions and the effect of $\delta R2_b$ is to reduce the
bare transition temperature  $T_{c2}^0$
(where $\alpha_0(T-T_{c1}) + \varepsilon_{0_{b}}(H=0) = 0)$
to $T_{c2}$, the observed lower transition temperature.
This is the precise analog of the critical temperature
for $M_z$ in Eq.\ \ref{eq:magtc}.   Also for $\bbox{H}$ in the basal plane
and arbitrary field strength, the problem can be solved
\cite{joynt91}.  At
the tetracritical point, $\delta R2_b$
vanishes, so there is no effective field.
Near this point, the effective fields can be calculated,
and we will carry this calculation out in the next section.
$ [\alpha_0(T-T_{c1}) + \varepsilon_{0_b}- \delta R2_b]$
vanishes along a line in $H$--$T$ space.
This represents the second phase transition for this field
direction.  In fact all the functions involved are
continuous and the second phase
transition occurs for all directions of $\bbox{H}$ for
UPt$_3$.

\subsubsection*{Class 3(b): terms determining the behavior of
$C_{n_bk}$ for $n_b > 0 $ and $k \in L.$}
The interesting new feature that arises
for these terms is that we now have an effective field from both
the $(C1)_a$ and the $(C2)_b$.
There are more terms in
the effective free energy for the $(C3)_b$.
However, we shall not consider this in detail, since it
is evident that these $(C3)_b$ couple linearly to
to the $(C1)_a$ and therefore start their life at $H_{c2}(T)$
where they are proportional to $\delta^{3/2}$.

\subsubsection*{Class 4(b): terms determining the leading behavior of
$C_{nk'}$, $n > 0$, $k' \in L' $.}
Here the relationship of terms
$(C4)_b$ to $(C2)_b$ is the same as that
s-wave $(C3)$ to s-wave $(C1)$.  Thus there is
an effective free energy of the form:
\begin{equation}
F_{{\text{\em eff}}}((C4)_b) \sim  [\alpha_0(T-T_{c1}) +
\varepsilon_{n_b}] (C4)_b^2 + {\cal O} ((C2)_b^3 (C4)_b).
\end{equation}
Crossovers only are allowed for these terms.

\subsubsection*{Class 5(b): other periodicities.}
We may neglect these for the same reasons as s-wave class 5.

Let us now summarize the conclusions.
At $H_{c2}$ the $C_{0_ak}$ condense to form a hexagonal lattice
which we call $L$.  At the same time, a number of other
coefficients begin to grow (such as those in class 3(a) and 1(b)),
though more slowly than the $C_{0_ak}$.  Their growth means that the
shape of the vortices is temperature and field-dependent, but the symmetry of
the lattice $L$ is unchanged.  As the temperature
or field is further lowered, the  $C_{0_bk'}$ become unstable,
forming a lattice $L'$ which interpenetrates $L$.  This occurs by a
second-order
phase transition.   This process is summarized in Fig.\ \ref{fig:htdwave}.

\section{Phase diagram: field in the basal plane}
\label{basal}

We have now established the mathematical method for finding the
phase boundaries.
In this section we apply the method to the {\it quantitative}
construction of the phase diagram for the case when the
field is in the basal plane of the hexagonal UPt$_3$ crystal.    We begin
by writing down the free energy density for a hexagonal $E_1$ or $E_2$ system:
\begin{eqnarray}
\label{fulleng}
f &=& \alpha_0 (T-T_x) |\eta_x|^2 + \alpha_0 (T-T_y) |\eta_y|^2 +
    \beta_1 ( \bbox{\eta} \cdot \bbox{\eta}^*)^2 +
    \beta_2 | \bbox{\eta} \cdot \bbox{\eta}|^2  \\
  & & \mbox{} + \sum_{i,j=x,y} ( K_1 D_i \eta_j D_i^* \eta_j^* +
  K_2 D_i \eta_i D_j^* \eta_j^* + K_3 D_i \eta_j D_j^* \eta_i^*) +
  K_4 \sum_{i=x,y} |D_z \eta_i|^2  \nonumber\\
  & & \mbox{} + (\alpha_0 \epsilon \Delta T) ( \hbar c/2e )
  \sum_{i=x,y} (|D_i \eta_x|^2 -|D_i \eta_y|^2) \nonumber \\
  & & \mbox{} + a_z H_z^2 \bbox{\eta}\cdot\bbox{\eta}^*
  +a_x (H_x^2+H_y^2) \bbox{\eta}\cdot\bbox{\eta}^*
  +a_d |\bbox{H}\cdot\bbox{\eta}|^2.
\end{eqnarray}
Here $\bbox{\eta} = (\eta_x,\eta_y)$ is the two-component order parameter,
and $K_1$, $K_2$, $K_3$, $K_4$, $\alpha_0$, $\beta_1$, $\beta_2$, $a_x$,
$a_z$, $a_d$ and $\epsilon$ are constants.  The coupling
of the staggered magnetization to $\bbox{\eta}$ is responsible for the
temperature splitting $\Delta T = T_x - T_y$.  The terms quadratic in $H$
are Pauli limiting terms.  They arise due to the reduction of the spin
susceptibility in the singlet superconducting state.  The effect of these
terms on the phase diagram and a physical explanation for the relative
sizes which we obtain for the various $a$ coefficients will be given in
Sec.\ \ref{magnetic}.
The phase diagram for the case of the field in the basal plane in our
theory has been considered previously by Joynt \cite{joynt91}, but we now
wish to consider it in more detail and compare our results to experimental
data.  Similar calculations have also been performed in other
models. \cite{luk'yanchuk,garg3}

It has been suggested that
the spatial variation of $\bbox{\eta}$ along the field direction
needs to be considered in the calculation of the upper
critical field. \cite{garg} To show that this does not occur,
we have computed the eigenvalues for the quadratic part of Eq.\ \ref{fulleng}
as a function of $p^2$, the wavevector along the
field direction.  The coefficient of $p^2$ is positive,
meaning that $\bbox{\eta}$ is uniform along the direction of the
field, unless
$(K_2+K_3)/K_1 > 3.126$.  As we shall see below,
this is certainly larger than
any value which can fit the upper critical field data.
In fact the ratio is roughly unity.
Let us choose a coordinate system such that $\bbox{H}=H\bbox{\hat{x}}$.
In this system, the result is that we can minimize
any terms in the free energy density containing $D_x$ by
setting them to zero.
Our free energy density is then
\begin{eqnarray}
\label{traneng}
f &=& \alpha_0 (T-T_y) |\eta_y|^2 + (\beta_1 + \beta_2)|\eta_y|^4 +
K_4|D_z \eta_y|^2+K^{\prime}_{123}|D_y \eta_y|^2 \\
  & & \mbox{} + \alpha_0 (T-T_x) |\eta_x|^2 + (\beta_1 + \beta_2)|\eta_x|^4 +
K_4|D_z \eta_x|^2 +K^{\prime}_1|D_y \eta_x|^2  \nonumber \\
  & & \mbox{} + 2 \beta_1 |\eta_x|^2|\eta_y|^2 + \beta_2({\eta_x^*}^2\eta_y^2
  + \mbox{C.C.}) \nonumber
\end{eqnarray}
In this equation $K_{123} = K_1 + K_2 + K_3$,
$K^{\prime}_{123}=K_{123}-(\alpha_0\epsilon\Delta T) (\hbar c/2e)$, and
$K^{\prime}_1=K_1+(\alpha_0 \epsilon \Delta T) (\hbar c/2e)$.

The upper critical fields of the
separate components $\eta_x$ ($H_{c2x}$) and $\eta_y$ ($H_{c2y}$) are now
easy to calculate, though the equations for the phase boundary are
more conveniently expressed in terms of the
inverse functions $T_{c2x}(H)$ and $T_{c2y}(H)$.  We obtain
\begin{eqnarray}
T_{c2x} & = &  -((a_x+a_d)/\alpha_0)H^2 + H/S_{c2x} + T_x \\
T_{c2y} & = &  -(a_x/\alpha_0)H^2 + H/S_{c2y} + T_y
\end{eqnarray}
where $S_{c2x} = -(\hbar c/2e)(\alpha_0 / \sqrt{K_4K^{\prime}_1})$ and
$S_{c2y} = -(\hbar c/2e)(\alpha_0/\sqrt{K_4K^{\prime}_{123}})$.
$S_{c2x}$ and $S_{c2y}$ are the slopes of the respective
$H_{c2}$ curves at zero field.
For $a_x$ and $a_d$ small (as we assume)
and if $T_y > T_x$ and $K^{\prime}_{123} > K^{\prime}_1$
then the two upper
critical field curves cross.  The physical phase boundary is the greater of the
two
and we obtain the well-known kink in $H_{c2}$
for this direction of the field.  We now want to find expressions
for the two inner transition lines $H_y^*(T)$ and $H_x^*(T)$.

Consider a fixed temperature $T$ such that $H_{c2y}(T) > H_{c2x}(T)$ and ask
what happens as the field is reduced starting from a field $H > H_{c2y}(T)$. As
the
field is lowered below $H_{c2y}(T)$ we will have $\eta_x = 0$ but $\eta_y
\neq 0$. From the parts of the free energy involving only $\eta_y$ we
immediately see that $\eta_y$ will form a lattice.
However, the lattice  will be distorted from pure
hexagonal because $K_4 \neq K^{\prime}_{123}$
in general.  If we choose the gauge $\bbox{A} = -Hz\bbox{\hat{y}}$ we obtain
\begin{equation}
\eta_y = N_y \sum_n c_n \exp(inqy - (z-nq\ell^2)^2/(2r_y^2\ell^2)).
\end{equation}
In this equation
$r_y^2 = \sqrt{K_4/K_{123}'}$, $N_y$ is the (real) magnitude of $\eta_y$, and
$q = \sqrt{\sqrt{3} \pi}(r_y/\ell)$.  Finally, $c_n = 1$ if $n$ is even and
$c_n = i$
if $n$ is odd.  The lattice formed by $|\eta_y|$ is a centered rectangular
lattice, athough it is perhaps more clearly imagined as a triangular
lattice which has been ``stretched'' by the anisotropy.  The side of the
unit cell parallel to $\bbox{\hat{y}}$ has a length of $2 \pi / q$, while the
side parallel to $\bbox{\hat{z}}$ has a length $q\ell^2$.

Suppose now that we are at a temperature $T$ such that $H_{c2x}(T) >
H_{c2y}(T)$  and we lower the field starting from $H > H_{c2x}(T)$.  In
this case as the field is lowered we will first come to a region where
$\eta_x \neq 0$ and $\eta_y = 0$.  We may then find $\eta_x$ in precisely
the same way as we found $\eta_y$ above.  Using the same gauge as before we
have
\begin{equation}
\eta_x = N_x \sum_n c_n \exp(inqy - (z-nq\ell^2)^2/(2r_x^2\ell^2)).
\end{equation}
Here, however, we have $r_x^2 = \sqrt{K_4/K_1'}$ and
$q = \sqrt{\sqrt{3} \pi}(r_x/\ell)$.

Let us return to the first case where we lower the field at constant $T$
and $H_{c2y}(T) > H_{c2x}(T)$ and ask what happens as the field is
lowered below $H_{c2y}(T)$.  Eventually the
field will be low enough so that the free energy is minimized with both
$\eta_x \neq 0$ and $\eta_y \neq 0$.  The point where this occurs is
$H_x^*(T)$.  To calculate $H_x^*(T)$ formally we need to follow the
prescription of Sec.\ \ref{math}: substitute in the functional forms for
$\eta_x$ and $\eta_y$ into $f$, compute the free energy $F = \int \/f dV$,
and determine when the coefficient of the term quadratic in $N_x$ changes sign.

We must choose the functional form of $\eta_x$ very carefully in this
calculation for
three reasons. First, we must allow the singularies of the $\eta_x$ flux
lattice to be located at different points than the singularities of the
$\eta_y$ flux lattice while still insuring that $\eta_x$ has a form
appropriate to the gauge we have chosen.  Second, we need to allow $\eta_x$ and
$\eta_y$ to have different phases. Finally, since $\eta_x$ is arising in the
effective periodic potential formed by $\eta_y$, we know that $\eta_x$ will
have the same periodicities as $\eta_y$.  (This is an application of
Bloch's theorem \cite{joynt91}).  Hence we write
\begin{equation}
\eta_x = N_x e^{i\theta}\sum_n c_n \exp(i(nq+z_0/\ell^2)(y-y_0)
 - (z-z_0-nq\ell^2)^2/(2r_x^2\ell^2))
\end{equation}
with $q = \sqrt{\sqrt{3} \pi}(r_y/\ell)$.

After minimizing the free energy
with respect to the phase difference $\theta$ the free energy is
\begin{eqnarray}
F & = & \alpha_0 [T-T_{c2y}(H)]< \mid \eta_y \mid ^2 > + (\beta_1+\beta_2)
     < \mid \eta_y \mid ^4 > \\
  & & \mbox{} +  \alpha_0 [T-T_{c2x}(H)]< \mid \eta_x \mid ^2 >
     + (\beta_1+\beta_2) < \mid \eta_x \mid ^4 > \nonumber \\
 & & \mbox{} + 2 \beta_1 < \mid \eta_x \mid ^2 \mid \eta_y \mid ^2 >
  - 2 \beta_2 \mid < {\eta_x^*}^2 \eta_y^2 e^{2i \theta} > \mid \nonumber \\
 & = & < f_x > + < f_y > + 2 \beta_1 I_1 - 2 \beta_2 \mid I_2 \mid.
\end{eqnarray}
Here the angle brackets denote a spatial average ($< \ldots > = \int \ldots
dV$), $f_x$ ($f_y$) is the part of the free energy density that depends only on
$\eta_x$ ($\eta_y$), $I_1 \equiv < \mid \eta_x \mid ^2 \mid \eta_y \mid ^2 >$,
and $I_2 \equiv < {\eta_x^*}^2 \eta_y^2 e^{2i \theta} >$.  The inner transition
($H_x^*(T)$) occurs when the term quadratic in $\eta_x$ changes sign or when
\begin{equation}
\label{trancond}
\alpha_0[T-T_{c2x}(H_x^*)]< \mid \eta_x \mid ^2 > + 2 \beta_1 I_1 - 2 \beta_2
\mid I_2 \mid = 0.
\end{equation}
This equation contains $N_y^2$ which is determined by minimizing $< f_y >$
which gives us
\begin{equation}
\alpha_0[T-T_{c2y}(H_x^*)]< \mid \eta_y \mid ^2 > + 2(\beta_1 + \beta_2) < \mid
\eta_y \mid ^4 > = 0.
\end{equation}
We may now solve for $H_x^*(T)$. Once again, however, it is easier to express
the result in terms of the inverse function $T_x^*(H)$. We obtain
\begin{equation}
T_x^*(H) = -(a_x^*/\alpha_0)H^2 + H/S_x^* + T_{x_0}^*
\end{equation}
where
\begin{eqnarray}
\label{eq:qy}
a_x^* & = & \frac{(a_x+a_d)-Q_ya_x}{1-Q_y},  \\
S_x^* & = & \frac{S_{c2x}S_{c2y}(1-Q_y)}{S_{c2y}-Q_yS_{c2x}}, \\
T_{x_0}^* & = & \frac{T_x-Q_yT_y}{1-Q_y}.
\end{eqnarray}
Here
\begin{equation}
\label{eq:Q}
Q_y \equiv \frac{\beta_1 I_1 - \beta_2 \mid I_2 \mid}{(\beta_1 + \beta_2)
\beta_A < \mid \eta_x \mid ^2 > < \mid \eta_y \mid ^2 >}
\end{equation}
where $\beta_A$ is the Abrikosov lattice parameter: $\beta_A = < \mid \eta_y
\mid ^4 > / ( < \mid \eta_y \mid ^2 > ) ^2$.  Note that we may rewrite our
result as
\begin{equation}
\label{Tx*repell}
T_x^*(H) = T_{c2x}(H) - \left ( \frac{Q_y}{1-Q_y} \right )
[T_{c2y}(H)-T_{c2x}(H)].
\end{equation}
{}From this equation we see that the the inner transition line for $\eta_x$
is repelled away from the calculated upper critical field curve for
$\eta_x$ by an amount proportional to the separation between the calculated
upper critical field curves for $\eta_y$ and $\eta_x$.

For the opposite case, when we consider lowering the field starting from a
temperature $T$ such that $H_{c2x}(T) > H_{c2y}(T)$, the calculation proceeds
precisely as before.  The inner transition line, $T_y^*$, for this case is
related to the calculated outer transition lines as in Eq.\ \ref{Tx*repell} so
that
\begin{equation}
\label{Ty*repell}
T_y^*(H) = T_{c2y}(H) - \left ( \frac{Q_x}{1-Q_x} \right )
[T_{c2x}(H)-T_{c2y}(H)].
\end{equation}
Here $Q_x$ is given by Eq.\ \ref{eq:Q}. However, since $\eta_y$ is becoming
non-zero in the periodic potential formed by $\eta_x$ the periodicity of the
flux lattices is set by $\eta_x$. This means that for this calculation
$q = \sqrt{\sqrt{3} \pi}(r_x/\ell)$.

We now must calculate the integrals $I_1$ and $I_2$.  This is a straightforward
but tedious exercise and we omit the details.  We only
note here that the results do not depend on $r_x$ and $r_y$ separately but
only on the ratio $r_x/r_y$.  This is significant for curve-fitting
because although $r_x$
and $r_y$ cannot be obtained from the phase diagram when $\bbox{H}$ is in
the basal plane their ratio can be obtained through $(r_x/r_y)^2 =
S_{c2x}/S_{c2y}$.  This equation follows directly from the definitions of
the quantities involved.

We also note that while $I_1$ is non-zero for all
values of the offset vector $\bbox{v} = y_0\bbox{\hat{y}} +
z_0\bbox{\hat{z}}$, $I_2$ is zero unless $\bbox{v}$ is a flux lattice lattice
vector or one-half of a flux lattice vector \cite{garg3}.  In other words if
$\bbox{u_1} = (2 \pi /q)\bbox{\hat{y}}$ and $\bbox{u_2} = (\pi
/q)\bbox{\hat{y}} + q \ell^2 \bbox{\hat{z}}$ are the basis vectors of the flux
lattice then $I_2$ is zero unless $\bbox{v} = \frac{1}{2}(n\bbox{u_1} +
m\bbox{u_2})$, where $n$ and $m$ are integers.  By symmetry there are only
three distinct offset points in the Wigner-Seitz primitive unit cell of the
flux
lattice where $I_2$ is non-zero.  These are the d ($\bbox{v} = 0$), c
($\bbox{v} = \frac{1}{2}\bbox{u_1}$), and b ($\bbox{v} = \frac{1}{2}
\bbox{u_2}$) points, as shown in Fig.\ \ref{fig:lattice}.   Consequently, by
the
definition of the $Q$'s (Eq.\
\ref{eq:Q}), $Q_x$ and $Q_y$ will have their smallest values when $\bbox{v}$ is
at one of these points.  By the equations for the inner transition lines
(Eqs.\ \ref{Tx*repell} and \ \ref{Ty*repell}) when
$Q$ is at its smallest the inner transition line is closest to the outer
transition line.  Hence, the inner transition line which is actually observed
is the transition line which corresponds to the smallest value of $Q$.  Hence,
we see that the inner transition line must correspond to an offset vector which
is at one of points d, c, or b.

To determine which offset vector is favored and to fit our theory to the
experimentally observed phase
diagram we must calculate $Q_y$, and from it $T_x^*(H)$, and $Q_x$, and from it
$T_y^*(H)$.    Along with the parameters that can be obtained
by requiring that the outer transition lines fit the data ($T_x$, $S_{c2x}$,
$a_d$, $T_y$, $S_{c2y}$, $a_x$) we also need $\beta_2/\beta_1$ in order to
perform the calculation.  This ratio may be
determined in from the specific heat jumps at zero field at the
outer ($\Delta C_V(T_{y})$) and inner ($\Delta C_V(T_{x_0}^*)$) transitions
using
\begin{equation}
\frac{\beta_2}{\beta_1} = \frac{\Delta C_V(T_{x_0}^*)/T_{x_0}^*}
{\Delta C_V(T_{y})/T_{y}}-\mbox{\Large 1}.
\end{equation}
{}From data for the specific heat jumps \cite{hasselbach2} we obtain
$\beta_2/\beta_1 = 0.5$.

The phase diagram we obtain from our calculations is shown along with the
ultrasonic velocity data from Ref.\ \cite{adenwalla} in Fig.\
\ref{fig:Hperpc} along with the values of the parameters used to obtain it.
We find that for our fit to the phase diagram the offset vector is at the b
point.  The fit is very good for the outer transition lines and $H_x^*(T)$ but
poor for $H_y^*(T)$.  The problems fitting $H_y^*(T)$ are
not difficult to understand.  The inner transition lines are given by
equations such as Eq.\ \ref{Ty*repell}.  These equations state that the inner
transition lines are repelled from the continuation of the corresponding
outer transition line by an amount which is proportional (in the case of
Eq.\ \ref{Ty*repell})
to $ Q_x/(1-Q_x)$.  The quantity $Q_x$ has been calculated in the limit
of very small $\delta$, i.\,e.\, near the tetracritical point.
Unlike the other quantities calculated, however, $Q_x$ is expected to
have very strong nonlinearities.  The first term in $Q_x$ is proportional to
\begin{equation}
\label{eq:ratio}
\frac{< \mid \eta_x \mid ^2 \mid \eta_y \mid ^2 >}{< \mid \eta_x \mid ^2 >
< \mid \eta_y \mid ^2 >}.
\end{equation}
This quantity is considerably less than one when the separation of the
vortices is comparable to the core size at the tetracritical point.
We find $Q_x = 0.333$ (see below).
When $H < H_{c2}$, however, the core size quickly becomes smaller
than the separation and
$|\eta_x|^2$ and $ |\eta_y|^2$ are constant except in the
region of the cores, only a small fraction of the volume.
Inspection of Eq.\ \ref{eq:ratio} shows that $ Q_x \rightarrow 1$
and $Q_x/(1-Q_x)$ becomes large.  This magnifies the repulsion between
$H_{c2y}(T)$ and $H_y^*(T)$ and therefore the repulsion between the phase
boundaries $H_{c2x}(T)$ and $H_y^*(T)$.

This brings in an additional error:
the breakdown of the our approximation for the form of the order
parameter.  We assumed that both  $\eta_x$ and $\eta_y$ were formed by the
usual linear combination of lowest Landau levels.  This assumption is strictly
correct only at the tetracritical point where the inner and outer transition
lines meet.  It remains a reasonable assumption as long as the inner transition
line is not repelled too far away from the outer transition line.  These
problems are
not so serious for $H_x^*(T)$, because in this case the curve fits
smoothly to the zero-field point, which is exact.  There is no such additional
constraint for $H_y^*(T)$.  It is therefore necessary to incorporate some
nonlinear effects in the fit to this line.  The slope at the
tetracritical point itself is corectly given by the linear calculation.
Over the length of the line, however, we use a renormalized $\tilde Q_x$ given
by fitting the slope.  Both renormalized and unrenormalized
fits are given in Fig.\ \ref{fig:Hperpc}.

\section{Phase diagram: field along the c-axis}
\label{caxis}

In this section we wish to take the free energy, Eq.\ \ref{fulleng}, and use it
to compute the phase diagram when the field is along the c-axis (the
$z$-direction) of the crystal.  The procedure for finding the outer transition
line (the upper critical field curve) in our theory for arbitrary angles of the
field with the c-axis has been developed elsewhere. \cite{sundaram,zhito}
We briefly review the procedure.

To find the upper critical field at an arbitrary field direction one first
follows the Euler-Lagrange prescription and demands that variations in the
free energy $F$ with respect to each component of the order parameter
vanish.  This condition gives two G-L equations which for
purposes of finding $H_{c2}$ may be linearized.  The linearized G-L
equations may be viewed as a Schr\"{o}dinger equation for $\bbox{\eta}$.
This defines an effective hamiltonian which is a $2 \times 2$ matrix in the
components of $\bbox{\eta}$.  One then defines a new
coordinate system with one axis along the field and the other two axes
perpendicular to it.

It is easy to show that the component of the $\bbox{D}$
operator along the field ($D_1$) commutes with the components in the other two
directions. Hence $D_1$ commutes with the effective hamiltonian and we may
rewrite any terms containing $D_1\bbox{\eta}$ as $p_1\bbox{\eta}$ where $p_1$
is a c-number.  When the field is in the $z$-direction the only terms which
result from this substitution are terms proportional to $p_1^2$, which are
minimized by setting $p_1 = 0$ Therefore in this case, as in the less
obvious case when the field is in the basal plane,  one may simplify the
G-L equations by setting $D_1 \bbox{\eta} = 0$.   Since this procedure may
be done both when the field is in the $z$-direction and in the seemingly
least favorable case when the field is in the basal plane it is reasonable
to assume that it may be done for any angle the field makes with the
$z$-axis.

One then
defines raising and lowering operators  $D_{\pm} = \ell (r D_2 \pm i D_3 / r)
/\sqrt{2}$ and $\eta_{\pm} = (\eta_x \pm i \eta_y)/2$.  Here $r$ is a function
of the angle the field makes with the c-axis and is chosen to simplify the G-L
equations as much as possible.  One can then rewrite the G-L
equations in terms of these quantities and expand $\eta_+$ and $\eta_-$
in terms of the states $\mid n >:$
\begin{equation}
\eta_+ = \sum_n a_n \mid n > \hspace{0.4in} \eta_- = \sum_n b_n \mid n >.
\end{equation}
Here $D_+D_- \mid n > = n \mid n >$.
The states $\mid n >$ are quasi-Landau levels.  The problem then splits into
finding
the eigenvalues of an infinite tri-diagonal matrix.
{}From the lowest eigenvalue
one can then compute $H_{c2}$.

Finding the inner transition line near to the upper critical field
is also a linear
problem, as the analysis of Sec.\ \ref{math} demonstrated.
To find the line rigorously we would have to calculate the
effective free energy for all of the eigenfunctions due to the presence of
the eigenfunction with the lowest eigenvalue, as outlined in Sec.\
\ref{math}.  However, we have seen that the only transition line which is not
destroyed (that is, either converted to a crossover or repelled to
non-physical fields and temperatures) by the coupling to the lowest
eigenfunction is the line which originates at the inner transition temperature
and corrresponds to a flux lattice shifted from the flux lattice formed by the
lowest eigenfunction. The full effective field matrix therefore
contains levels which are pushed to unphysical fields (pushed up
to high energy in the quantum-mechanical analogy), or have small
magnitude $( \sim \delta^{3/2})$.  Thus, rather remarkably,  it will be a very
good
approximation to compute the inner transition line using only two
levels.  As in the case when the field is in the basal plane we then
have a correction to the bare inner transition line which is
proportional to the separation between the bare inner transition and the
outer transition in order to find the actual inner transition line.  Our
formula for the inner transition line $T_{\text{\em inner}}(H)$ in terms of
the bare inner transition line $T_{\text{\em bare}}(H)$ and the outer
transition line $T_{\text{\em outer}}(H)$ is then
\begin{equation}
T_{\text{\em inner}}(H)  =  T_{\text{\em bare}}(H) - g[T_{\text{\em outer}}(H)
-T_{\text{\em bare}}(H)].
\end{equation}
When the field was in the basal plane we were able to
compute the coupling constant $g$.  For the field along the c-axis
this computation, though straightforward in principle, is
exceedingly complicated.  Accordingly,
$g$ is found by fitting to the data.

A key feature of the bare inner transiton line comes to light upon
examining the matrix used to find it.  This matrix is
\begin{equation}
\left ( \begin{array}{cc}
(2K_1+K)-K' & -\alpha_0 \ell^2 \Delta T (2 \epsilon H - 1) \\
-\alpha_0 \ell^2 \Delta T (2 \epsilon H - 1) & (2K_1+K)+K'
\end{array} \right ).
\label{eq:matrix}
\end{equation}
Here $K = K_2 + K_3$ and $K' = K_2 - K_3$.
Note that the off-diagonal terms will vanish when $H = 1/2 \epsilon$.  This
means that if $K' \approx 0$, as is expected from particle-hole symmetry
\cite{p-h}, the two eigenvalues will be nearly degenerate at this $H$ and
the outer and bare inner transition lines will nearly touch. This
cancellation between the derivative terms in the free energy (Eq.\
\ref{fulleng}) proportional to $K_2$ and $K_3$ and the terms which couple
the staggered magnetization (through $\Delta T$) to the derivatives means we
obtain an {\em apparent} tetracritical point - the two lines come
close but do not quite touch. There are {\em only two} superconducting
phases when the field is in the $z$-direction or indeed for any
direction except in the basal plane.  The fact that $K'$
depends sensitively on the impurity density has interesting consequences.
The miminum separation
will depend on this density.  Unfortunately, the sharpness of these
transitions also depends on the impurity density.

The phase diagram we obtain from our calculations with the field in the
$z$-direction together with the values of
the parameters used to obtain it and the ultrasonic velocity data from
Ref.\ \cite{adenwalla} is shown in Fig.\ \ref{fig:Hparc}. Note that we use the
same parameters as for our fit for when the field is in the basal plane
along with some additional parameters.  As was the case when the field was
in the basal plane the fit is very good except for the high-field,
low-temperature part of the inner phase boundary where the linear theory is
expected to break down.  This happens because
of the renormalizations discussed in the previous section.

A virtue of the theory given here is that a striking difference between the
phase diagrams for the two directions of the field receives an
explanation.  The upper critical field curve is smoother for
field along the c-axis, and the inner transition line
is much smoother.  This can now be seen to result from
the `hybridization'  of the two curves for this case,
the presence of the off-diagonal matrix elements in
Eq.\ \ref{eq:matrix}.  This is absent for the other field direction,
when the two components decouple.

\section{Magnetic properties of UP\lowercase{t}$_3$}
\label{magnetic}

In this section we wish to discuss the origin, effect, and relative sizes
of the Pauli limiting terms in the free energy density.
These are the terms proportional to $H^2 \eta^2$ in Eq.\ \ref{fulleng}.
In order to do this, we require some preliminary
background about magnetic properties of the normal state.
The first fact to appreciate is that the magnetic susceptibility $ \chi_{ij}$
of the normal state is enhanced by roughly the same factor as the
mass.  Because  $ \chi_{ij}$ is large, the Pauli limiting effect of the field
on superconductivity is likely to be appreciable.
The second important point is that the susceptibility is
anisotropic, and the temperature dependences of the components
are different.  This is clear from the plots of
the susceptibilities $\chi_{xx}(T)$ and $\chi_{zz}(T)$ in Fig.\ \ref{fig:chi}.
$\chi_{xx}(T) > \chi_{zz}(T)$ at all temperatures. \cite{frings}
At high $T$, both functions take on the local moment form $\chi \sim 1/T$,
while each goes to a finite constant, characteristic of
Pauli or van Vleck behavior, at low $T$.  In addition,
$\chi_{xx}(T)$ has an anomaly around 15 K.

Let us first take a theoretical approach to
understanding the anisotropy in $\chi_{ij}$.
Our basic assumption is that
UPt$_3$ is a Fermi liquid at temperatures just above the
critical temperature.  Then
the starting point is the single-particle states calculated in
band theory, which account very well for the
Fermi surface. \cite{wang}  The
states near the Fermi surface are predominantly derived from
uranium 5f orbitals with $j=5/2$, as would be expected
from Hund's rules for an actinide system with
a 5f occupancy near 2.
In the isolated atom, the $j=5/2$ level is
6-fold degenerate.  In the hexagonal crystal field,
there is an effective Hamiltonian  at the $\Gamma$ point
which splits the six-fold
degenerate state into three doublets at the $\Gamma$ point:
$j_z = \pm 5/2$, $j_z = \pm 3/2$, and $j_z = \pm 1/2$.
This means that UPt$_3$
is likely to be an example of a system in which the magnetism is
Van Vleck-like in the plane and Pauli-like along the c-axis,
which is expected to be a general feature of hexagonal
U-based systems. \cite{dan}.

Let us briefly review the reasons for this expectation.
If we apply a magnetic field, there will be both
a Pauli (intraband) and a Van Vleck (interband) contribution to the
susceptibility.  The former is of order $(g_{\text{\em eff}}
 \mu_B)^2 N(\varepsilon_F)$,
while the latter is of order $(g_{\text{\em eff}} \mu_B)^2/ |B_h| $.
Here $g_{\text{\em eff}}$ is
an effective g-factor for the coupling of the field to the total
angular momentum of the band or bands involved.  It is a dimensionless
number of order unity.  The Land\'{e} factor for $\ell$=3, s=1/2 ,
and $j$=5/2 is 6/7.   $B_h$ is the separation between the bands and
$N(\varepsilon_F)$ is the density of states at the  Fermi energy.
The Van Vleck susceptibility is given by
\begin{equation}
\chi_{ii}=2n\mu_{B}^{2} \sum_{\alpha,\beta}
\frac
{|<\alpha|L_{i}+2 S_{i}|\beta>|^2}
{E_{\beta}-E_{\alpha}}
f_{\alpha}(1-f_{\beta}).
\end{equation}
Here $f_\alpha$, $f_\beta$, $E_\alpha$, $E_\beta$ are occupation factors
and energies of the states $\alpha$ and $\beta$.
In view of the greater
multiplicity of the interband transitions, we
expect the Van Vleck susceptibility to be very important - indeed it
very likely dominates the total.  A band calculation which
explicitly computes the two components reckons the Pauli contribution
at 15-20\%, \cite{mike} in rough agreement with this multiplicity argument.

If $\bbox{H}$ is along the c-axis, then
the relevant matrix element (with $\hbar = 1$) is:
\begin{equation}
|<\alpha| L_z+2 S_z|\beta>|^2
=(36/49)j_z^2 \delta_{\alpha,\beta}.
\end{equation}
At the $\Gamma$ point, states of different $j_z$
do not mix and
the perturbation introduced by $\bbox{H}$ is diagonal.
The occupation factors then imply that the Van Vleck susceptibility is
zero for this direction.  If $\bbox{H}$ is in the $x$-direction,
the corresponding expression for the square of the matrix element is
\begin{equation}
|<\alpha| L_x+2 S_x|\beta>|^2 =
(36/49)(5/2-j_z)(5/2+j_z+1)
\end{equation}
if the states $\alpha$ and $\beta$ differ by one unit of $j_z$
and is zero otherwise.  The Van Vleck susceptibility comes from
four distinct pairs of states:($j_z=-5/2,-3/2$), ($-3/2,-1/2$), ($1/2,3/2$) and
($3/2,5/2$), whenever one of the pair is occupied and the other unoccupied.
The Pauli contribution to $\chi_{xx}$, on the other hand, comes only from
the pair ($-1/2,1/2$) when this state is occupied.
A sheet of the Fermi surface will have an isotropic
partial Pauli susceptibility ($\chi_{zz}^P/\chi_{xx}^P \approx 1$)
if different $j_z$ values are well mixed in the
wavefunction, but will be anisotropic otherwise:
$j_z$=1/2 implies $\chi_{zz}^P/\chi_{xx}^P << 1$,
and $j_z$=3/2 or $j_z$=5/2 implies $\chi_{zz}^P/\chi_{xx}^P >> 1$.
As we shall see below, it is the anisotropy
of the Pauli contribution which is critical for
understanding the phase diagram.  This
means that the central question is: what is the
$j_z$ content of the Fermi surface, and how much
mixing of different $j_z$'s is there?
Band calculations give a clear answer to this question.
They show that the parts of the Fermi surface near the
$\Gamma$ point and K point are predominantly  $j_z$=3/2 or 5/2,
\cite{zwicknagl,andersen} while the parts near the A point are well mixed.
This is illustrated in Fig.\ \ref{fig:band}.
Hence we expect a contribution to the Pauli susceptibility which satisfies
$\chi_{zz}^P/\chi_{xx}^P \gg 1$ from the parts near
$\Gamma$ and K, representing roughly half the
total density of states at the Fermi surface, and a contribution satisfying
$\chi_{zz}^P/\chi_{xx}^P \approx 1$ for the
rest of the Fermi surface.
In treatments which go beyond band theory to discuss many-body
renormalizations, it is found that the Pauli and Van Vleck parts are enhanced
by
similar factors \cite{fuchun}.

Summing up these theoretical considerations, the magnetic
susceptibility of UPt$_3$ is likely to be dominated by
interband (Van Vleck) contributions.
This is particularly true for $\chi_{xx}$,
which means that the anisotropy in the oberved susceptibility
($\chi_{xx} > \chi_{zz}$) most likely stems from
interband contributions.  The Pauli susceptibility,
on the other hand, is more likely to satisfy the opposite
inequality $\chi_{xx}^P < \chi_{zz}^P$

Experimentally, it is not easy to distinguish
the Pauli and Van Vleck contributions to the susceptibility.
The most straightforward way, in principle, is to
measure the imaginary part of the susceptibility
with neutron scattering.  The Van Vleck contribution
has a gap at low frequencies, while the Pauli part does not.
For the present case, however, we also need to distinguish the
different components of the susceptibility tensor.
This means that polarized beam experiments are required,
with the associated lower counting rates.  Finally,
we are interested here in the uniform susceptibility,
which means small-angle scattering.  Thus this
definitive experiment may be difficult to perform.

A more indirect but still informative test arises
from the observation that the Pauli susceptibility
depends on the density of states at the Fermi energy whereas the
Van Vleck susceptibility depends on a joint density of states.
The Pauli part is therefore directly comparable to $C_V/T$,
where $C_V$ is the specific heat.
In this regard the peak in $\chi_{xx}(T)$
at T=15 K \cite{frings}  (see Fig.\ \ref{fig:chi})
is of interest.
This peak is absent in the
smooth curve for $\chi_{zz}(T)$, and in the
the specific heat $C_V(T)$, \cite{devisser}
This is consistent with the idea that the physical origins of
$\chi_{zz}$ and $\chi_{xx}$ are different, and that the
density of states at the Fermi level
largely determines $\chi_{zz}$ but not $\chi_{xx}$.
Thus experiments, to the exent that we have them,
confirm the theoretical picture.

The importance of these considerations for the superconducting
state is simple. \cite{cox}
Superconductivity affects the Pauli susceptibility
in a drastic fashion.  For a singlet state such as $E_{1g}$,
the Pauli term $\chi_{ij}^P(T)$
is reduced to zero at zero temperature because
it takes a finite amount of energy to break a pair and magnetize
the system.  Superconductivity should have no effect at all
on the Van Vleck term, and conversely.  The difference in
free energies between the
normal and superconducting states in a field is
\begin{equation}
F_{magnetic}=-\frac{1}{2} \sum_{ij} \Delta\chi^P_{ij} H_{i} H_{j}.
\end{equation}
Here $\Delta \chi_{ij}^P = \chi_{ij}^S - \chi_{ij}^N$ where $\chi_{ij}^S$
and $\chi_{ij}^N$ are the Pauli susceptibilities in the superconducting and
normal states, respectively.  Just below the superconducting transition we
know that the change in the susceptibility is quadratic in $\bbox{\eta}$.
Hence we add to the usual superconducting free energy the last three terms
of Eq.\ \ref{fulleng} which are quadratic in both $\bbox{\eta}$ and $H$.

{}From the arguments above we expect that $a_x$ and $a_d$ will be smaller
than $a_z$ since we anticipate that $\chi_{xx}^P < \chi_{zz}^P$.  From our
fits to the phase diagrams for the two directions of the field we find that
$a_z$ is slightly more than twice $a_x+a_d$, in agreement with the physical
picture of the susceptibility.  The differences in the sizes of the $a$
terms affect what happens to the upper critical field curves for the two
directions of the field at high fields.  At high fields the Pauli limiting
terms in the free energy, which are proportional to $H^2$, dominate over
the rest of the free energy, which gives a contribution to $H_{c2}$
proportional to $H$.  Because $a_z > a_x+a_d$, $H_{c2}$ when the field is
along the c-axis curves down more than $H_{c2}$ when the field is in the
basal plane.  Consequently, the two upper critical field curves for the two
directions cross.  This crossing is shown in Fig.\ \ref{fig:crossing}.
We have therefore shown that the objection to the $E_{1g}$ model on the
grounds that it cannot explain the crossing of the upper critical field
curves is invalid.

\section{Pressure effects}
\label{press}
We have offered a comprehensive description of the phase diagram of
UPt$_3$ in the $H-T$ plane.  However, because of the rather large number of
parameters in the Ginzburg-Landau free energy, this analysis is
not yet sufficient to distinguish the $E_{1g}$ picture from competing pictures
such as the $E_{2u}$ picture and mixed representation pictures.
Consideration of pressure effects will allow us to do this.  We will
show that only $E_{1g}$ is consistent with these experiments.
The analysis in this section is an elaboration of earlier work.\ \cite{joynt93}
It is somewhat surprising that pressure experiments are so crucial
for understanding the symmetry of the order parameter.  Under normal
circumstances,
accessible pressures have only a small effect on superconducting
parameters and qualitative conclusions are difficult to draw.  In the present
case, however, moderate pressures destroy antiferromagnetism, which
restores the full hexagonal symmetry of the crystal structure.  It is
this singularly fortunate circumstance which makes pressure
such a very powerful tool in unraveling the order parameter symmetry.

Qualitatively, the facts are these.  The magnetization disappears at
a critical pressure of about 3 kbar.  The splitting in $T_c$
also disappears at the same pressure.  This shows that it is
indeed the magnetization which splits the transition, as
originally predicted \cite{joynt88}.  The coincidence of the
pressures at which these events take place rules out
mixed representation theories such as the A-B theory \cite{garg}.
In such theories the original splitting is due to an accidental
degeneracy and is not related to the magnetization.

Our aim is to understand quantitatively the phases of UPt$_3$
in the entire ($H, P, T$) space.  However,
in order to understand the restoration of crystal symmetry,
we first focus on the ($H=0,P,T$) plane, so that complications due to
the gradient terms can be treated separately.  The expression for the
free energy density of the coupled magnetic-superconducting system is
then $f=f_{S}+f_{M}+f_{SM}$, where
\begin{equation}
f_{M}=\alpha_{M}(P,T)M^{2}+\beta_{M}M^{4}
\label{eq:fm}
\end{equation}
\begin{equation}
f_{S}=\alpha_{S}(P,T)\bbox{\eta}\cdot\bbox{\eta}^{*}+
\beta_{1}(\bbox{\eta}\cdot\bbox{\eta}^{*})^{2}+
\beta_{2}|\bbox{\eta}\cdot\bbox{\eta}|^{2}
\label{eq:fs}
\end{equation}
\begin{equation}
f_{SM}=b|\bbox{M}\cdot\bbox{\eta}|^{2}+b'M^{2}\bbox{\eta}\cdot\bbox{\eta}^{*}
\label{eq:fsm}
\end{equation}
We have assumed, as is conventional,
that the presure dependence of fourth-order coefficients is weak and
can be neglected.

$f_M$, the magnetic part of the free energy, entirely determines
the behavior of the magnetization above $T_{c1}$. (Recall that $T_{c1}$ is
the higher of the two observed transition temperatures.) The experimental data
from neutron scattering measurements of $M^2$ (proportional to
the magnetic Bragg scattering
at the $(1,\frac{1}{2},0)$ point) are sufficient to determine the parameters.
At $P=0$ and $T>T_{c1}=0.5$ K, $M^{2}$ is a linear function of $T_{N}-T$,
where $T_{N}=5$ K is the Neel temperature.\ \cite{aeppli,hayden}  One finds
$\alpha_{M}(P=0,T)/\beta_{M}= (1.6\times10^{-4}\mu_{B}^{2}/$K$)(T-T_{N}$).

As to the $P$ dependence, it is found that $T_{N}$ is nearly independent of
pressure from
$P=0$ to $P=2$ kbar and that $M^{2}\sim (P_{N}-P)$ for $T<2$ K \cite{hayden},
where $P_{N}\approx 3$ kbar is the critical pressure at which
magnetism disappears.  From the point of view of this paper, which
concentrates on the superconducting regime $T<T_{c1}$, we may
therefore take $\alpha_{M}=\alpha_{M}^{0}(P-P_{N})(T-T_{N})
\approx-\alpha_{M}^{0}T_{N}(P-P_{N})$, where
$\alpha_{M}^{0}/\beta_{M}=5.3\times10^{-5}\mu_{B}^{2}$/K-kbar. Note that
this value and the coefficient of the expression for $\alpha_{M}(P=0,T)/
\beta_{M}$ have been corrected from an earlier paper written by one of us
(Joynt) \cite{joynt93}.

The pressure dependence of $\alpha_{S}(P,T)$ may be obtained
if we assume that $\alpha_{S}(P,T) = \alpha_{ST}(T-T_c^0) +  \alpha_{SP}P$,
so that $\alpha_{S}(P,T)$ is a linear function of $P$.  $\alpha_{ST}(T)$
is the zero pressure value of $\alpha_{S}(P,T)$  which has already
been determined. For $P > P_N$, $M=0$ and the pressure dependence of
$T_c$ is entirely due to the coefficient $\alpha_{SP}$.
Since $dT_c/dP = -11$ mK/kbar in this region \cite{trappmann},
we find $\alpha_{SP} =  \alpha_{ST}$(11 mK/kbar).

At $P=0$ and $T<T_{c1}$, there is a
competition between the purely magnetic terms
and the coupling term $f_{SM}$.  Because
$\bbox{\eta}\cdot\bbox{\eta}^{*}\sim T_{c1}-T$ for $T<T_{c1}$
and $\bbox{\eta}=0$ for $T>T_{c1}$, the coupling term predicts
that $\bbox{M}$ should have a kink at $T_{c1}$.
The magnitude of the kink may easily be computed using
Eqs.
\ref{eq:fm},
\ref{eq:fs},
\ref{eq:fsm}.
Differentiation leads to two linear equations for $M$ and $\bbox{\eta}=\eta
\bbox{\hat{x}}$

\begin{equation}
\eta^2 = [\alpha_S(T_c^0-T) - (b+b') M^2] / 2 (\beta_1 + \beta_2)
\end{equation}

\begin{equation}
M^2 = [\alpha_M(T_N-T) - (b+b') \eta^2] / 2 \beta_M,
\end{equation}
which give the behavior of the order parameters below
$T_{c1}$.  Above $T_{c1}$ we have simply

\begin{equation}
M^2 = \alpha_M(T_N-T)/ 2 \beta_M
\end{equation}
The slope is discontinuous at $T_{c1}$:

\begin{equation}
 \left [ 1 - \frac{(b+b')^2}{4 \beta_M (\beta_1
+ \beta_2)} \right ] \left.\frac{dM^2}{dT}\right|_{T<T_{c1}} =
\left.\frac{dM^2}{dT}\right|_{T>T_{c1}} + \frac{\alpha_{ST}(b+b')}
{4 \beta_M (\beta_1 + \beta_2)}.
\end{equation}
If we take the approximation that the coupling $(b+b')$
is small, then we may write the discontinuity as
\begin{equation}
\Delta \frac{dM^2}{dT} = - \frac{\alpha_{ST}(b+b')}
{4 \beta_M (\beta_1 + \beta_2)}.
\end{equation}
In these formulas $\bbox{\eta}$ is assumed to be parallel to $\bbox{M}$.
If these two vectors are perpendicular, then $b$ should appear instead of
$(b + b')$.  The kink is observed experimentally, \cite {aeppli}
which again confirms that the splitting of the superconducting
transition is due to $f_{SM}$.
These formulas assume that there is only one component of $\bbox{M}$,
contrary to the idea of Blount {\em et al.} \cite{varma}
that the moment rotates
at $T_{c1}$.  Recent experiments have indeed ruled out the possibility of
rotation \cite{luke}.

We now wish
to calculate the phase diagram at finite pressures,
assuming that the only pressure dependence comes from
$\alpha_S$ and $\alpha_M$.  All other parameters are
taken to have their zero pressure values.  The only dependence on pressure
in our theory of the phase diagram is through the quantity $\Delta T$.  We
calculate $\Delta T$ at various pressures by taking $T_x(P=0)$ and
$T_y(P=0)$ from our zero pressure fit, $dT_y/dP$ (recall $T_y > T_x$), $P_N$
and $dT_c/dP$ (for $P > P_N$) from experimental data \cite{trappmann}.
{}From $T_x(P=0)$ and $P_N$ we can then find $dT_x/dP$.
In Fig. \ref{E1gpressure} we plot the phase diagram in the
$H-T$ plane at various pressures for $\bbox{H}$
in the basal plane using the renormalized $Q_x$ (see section \ref{basal}).
The behavior with pressure is easily understood qualitatively.  The main effect
of pressure is to close up the splitting of the zero-field
critical temperatures.  Thus the tetracritical point
moves down toward the $T$-axis and disappears, as does the
A (low field, high temperature) phase.  Thus the C (high field, low
temperature) - B (low field, low temperature) phase boundary is very sensitive
to pressure as it ends at the tetracritical point.
This is observed experimentally.\ \cite{vandijk}
On the other hand, the N (normal state) - C boundary (upper part of the
$H_{c2}$ curve) is not very sensitive to pressure.
Again, the agreement between theory and experiment is
very satisfactory.
It is difficult to compare these predictions with
the threee-dimensional phase diagram of Boukhny {\em et al.\ } \cite{boukhny}
quantitatively.
The pressure dependence of the critical temperatures
given by these authors is not in good agreement
with that of Trappmann {\em et al.\ } \cite{trappmann},
which we used in plotting the figures.
The behavior of the boundaries is quite sensitive to this
dependence.  Nevertheless
there appears to be very satisfactory qualitative agreement
between theory and experiment, with one exception.
The experiment shows that there is an additional phase boundary
in the $P$-$T$ plane when $P>P_N$.  This cannot be a pure
superconducting transition in a two-component theory.
We believe this to be a mixed magnetic-superconducting transition,
so that this boundary is essentially an extension of the magnetic
phase boundary.  The signal in the sound velocity is very
small.  It may be larger than in the normal phase because of the
coupling to the superconducting order parameter which is serving as
a secondary order parameter in the transition.

Let us compare this behavior to the behavior of the phase boundaries
in the $E_{2u}$ theory in which $K_2 \approx K_3 \approx 0$.
The best fit with this constraint is given in Fig.\ \ref{E2upressure}.
This picture is in qualitative disagreement with
experiment.  Again, the qualitative reason for this is easily
understood.  In the $E_{2u}$ theory, the difference in slope
between the $H_{c2}$ curves for $\eta_x$ and $\eta_y$
is due only to their differing energies in the presence
of the magnetization.  Once $P > P_N$, this difference is
gone and the two components have identical free energies and
identical slopes.  The $E_{2u}$ theory says that the N-A and A-B boundaries
must move together, not apart, under the influence of
pressure.  This is in conflict with experiment.

\section{Conclusion}
\label{ending}

The Ginzburg-Landau theory is a very powerful tool in the
physics of unconventional superconductivity.  We have
pushed the theory to obtain as much information as possible
about the phase diagram.  Mathematical difficulties arise when
a magnetic field is applied, a circumstance which has made the
theory of the phase diagram of UPt$_3$ proceed more slowly than might
have been expected.  The method of classifying terms according to
their behavior in the effective field appears to have solved the
linear problem in principle, though explicit calculations for
a general direction of the field still appear daunting.  We have
limited our treatment to the two high symmetry field directions.
Most experiments are also limited to these directions.

Consistent application of the method, taking into account
the Pauli-limiting effect, gives very good agreement between
theory and experiment for the $E_{1g}$ theory.  It would be desirable,
however, to have an explicit calculation of the nonlinear
renormalization factors entering the repulsion of the phase boundaries;
obtaining this by a fit, as done here, is not truly satisfactory
from the theorist's point of view.
Thr peculiar phenomenon of the $H_{c2}$ crossing is interpreted here
as arising from an interplay of intraband Pauli magnetism
and interband Van Vleck magnetism.  While the picture
of the anisotropic susceptibility which emerges is a natural one,
it would be good to have some independent confirmation of it.

The surprise of the past several years is that pressure experiments
have been able to play a critical role in sorting out the
nature of the order parameter.  They have demonstrated that it is the
magnetism which splits the critical temperature.  Above the critical
pressure, the hexagonal symmetry is restored.  Experiments above
this pressure have shown that there are still two phase
transitions as a function of field - this means that the field
direction itself couples to the internal degrees of freedom
in the two-component order parameter.  This only occurs in the $E_{1g}$
picture, which appears to be the only choice fully consistent with
all experiments.

We would like to acknowledge useful discussions and correspondence with
M. Norman, A. Garg, and particularly D. Cox.  This work was supported by
the National Science Foundation through grant no. DMR-9214739.

\begin{figure}
\caption{Consequences of the effective field in the s-wave case.
(a) Eigenvalue curves for the original lattice with $k \in L.$
The lines are the solutions to Eq.\ \protect
\ref{eq:hc2}.  However, all transitions
are suppressed by the effective field except the original $H_{c2}$ line.
In the case of arrows, the repulsion
of the boundary to unphysical values of $H$ and $T$ takes place. In the case
of cross-hatching, the transition is converted to a crossover. The numbers in
parentheses denote the class of the line.
(b) Curves for the new lattice with $k \in L'$.
As in (a), except that the new lattice
interpenetrates the old one. All lines are repelled by the effective field,
and the $C_{nk}$ corresponding to these boundaries never become nonzero.}
\label{fig:htswave}
\end{figure}

\begin{figure}
\caption{Consequences of the effective field in the d-wave case.
(a) Eigenvalue curves for the original lattice with $k \in L$
setting the eigenvalues of the quadratic form in Eq.\ \protect \ref{eq:df}
equal to zero, neglecting the fourth-order terms.
Most transitions are converted to crossovers
by the effective field except the original $H_{c2}$ line. The numbers in
parentheses denote the class of the line.
(b) New lattice with $k \in L'$.  As in (a), but the new lattice
interpenetrates the old one.
Transitions corresponding to $C_{0_a}$ and $C_{0_b}$ are repelled
only a short distance (single arrows).  The dashed lines show the final
positions of these boundaries after taking into account the effective
field.  A single internal transition line remains.}
\label{fig:htdwave}
\end{figure}

\begin{figure}
\caption{The flux lattice for $\eta_y$.  The filled circles indicate the
singularities in the flux lattice. The Wigner-Seitz primitive unit cell is
indicated by the dashed line.  Its basis vectors are also shown. The integral
$I_2$ ($\mid I_2 \mid = \mid < (\eta_x^*)^2 \eta_y^2> \mid$) is non-zero only
if the singularities in the $\eta_x$ flux lattice are located at the b (open
square), c (open oval) or d point (x), or one of the symmetrically equivalent
points.  These points for the b (c) point are the filled squares (filled oval).
The axes are $\bar{z} = z/\ell$ ($\parallel$ c-axis) and $\bar{y} = y/\ell$.
The diagram is drawn to scale using the values of the parameters we obtain
through our fits.  The flux lattice for $\eta_x$ is identical except for
greater elongation in the $\bar{z}$ direction.}
\label{fig:lattice}
\end{figure}

\begin{figure}
\caption{Phase diagram when the field is in the basal plane.  The data points
are from ultrasonic velocity measurements and are taken from Ref.\
\protect \cite{adenwalla}, Fig.\ 3.  The solid lines are our fit without
the renormalization discussed in the text.  The dashed line coresponds to a
renormalization of $Q_x$ by 2.67 or $\tilde{Q}_x = 2.67Q_x$.  The constants
used to make these graphs are $T_x = 0.458$ K, $T_y = 0.504$ K, $S_{c2x} =
-9.26$ T/K, $S_{c2y} = -4.39$ T/K, $a_x/\alpha_0 = 0.0138$ K/T$^2$, and
$a_d/\alpha_0 = 0.0193$ K/T$^2$}
\label{fig:Hperpc}
\end{figure}

\begin{figure}
\caption{Phase diagram when the field is in the $z$-direction. The lines
are the theoretical fits to $H_{c2}$ (solid line) and the inner transition
(dashed line).  The data points are from ultrasonic velocity measurements
and are taken from Ref.\ \protect \cite{adenwalla}, Fig.\ 3.  In addition
to the constants used to make Fig. \protect \ref{fig:Hperpc} we have used
$\alpha_0/K_1=5.6\times 10^{12}$ K$^{-1}$cm$^{-2}$, $\epsilon=5.26\times
10^{-5}
$ G$^{-1}$, $a_z/\alpha_0=6.3\times 10^{-10}$ K/G$^2$, and $(K_2-K_3)/K_1=0.1$.
We use $(K_2+K_3)/K_1=(S_{c2x}/S_{c2y})^2-1+\hbar c\alpha_0\epsilon \Delta T
[(S_{c2x}/S_{c2y})^2+1)]/(2eK_1)$ and $K_4/K_1=[(1+\alpha_0\epsilon\Delta
T/K_1)[2
eK_1S_{c2x}/(\hbar c\alpha_0)]^2]^{-1}$ to obtain $(K_2 + K_3)/K_1=1.0$ and
$K_4/K_1=7.20$.}
\label{fig:Hparc}
\end{figure}

\begin{figure}
\caption{Susceptibility of UPt$_3$ for fields oriented along the crystal's
a-axis (circles), b-axis (triangles), and c-axis (squares).  The graph is
taken from Ref.\ \protect \cite{devisser}, Fig.\ 2.1.}
\label{fig:chi}
\end{figure}

\begin{figure}
\caption{Fermi surface of UPt$_3$ from  Ref.\ \protect \cite{andersen}, Fig.\
5.
The Fermi surface shown was calculated in the local density approximation using
the Dirac-relativistic linear muffin-tin orbital method.  The stripes show
the $j_z$ content of the Fermi surface: dotted for $|j_z|=1/2$,
right-hatched (///) for $|j_z|=3/2$, and left-hatched for
$|j_z|=5/2$. The $|j_z|=7/2$ component is small over the entire Fermi surface.
Note that the parts of the Fermi surface around the $\Gamma$ and K points are
predominantly $|j_z|=3/2$ or 5/2, while the parts around the A point have
well-mixed $|j_z|$'s. Fig.\ 2(b) of Ref.\ \protect \cite{zwicknagl} is
similar.}
\label{fig:band}
\end{figure}

\begin{figure}
\caption{The crossing of the $H_{c2}$ line when the field is the
basal plane (solid line) and the $H_{c2}$ line when the field is in the
$z$-direction (dashed line).  The data points are from ultrasonic velocity
measurements and are taken from Ref.\ \protect \cite{adenwalla}, Fig.\ 3. The
diamonds are for the case when the field in the basal plane
($\protect \bbox{H} \parallel $ ab)
and the crosses are for the case when the field is in the $z$-direction
($\protect \bbox{H} \parallel $ c).}
\label{fig:crossing}
\end{figure}

\begin{figure}
\caption{Pressure dependence of the phase diagram with the field in the basal
plane in the $E_{1g}$ model. The phase diagram is plotted at pressures ($P$)
of  (a) $P = 0$, (b) $P = P_N/2$ (1.85 kbar), (c) $P = P_N$ (3.7 kbar), and
(d) $P = (3/2)P_N$ (5.55 kbar).  Here $P_N$ is the pressure above which the
temperature splitting vanishes.  As discussed in the text the theoretical
inner transition line for temperatures below the tetracritical point
($H_y^*(T)$) has been renormalized.  The data points in (a) are taken
from Ref.\ \protect \cite{adenwalla}, Fig.\ 3.  The variation of the
transition temperatures with pressure is taken from Ref.\ \protect
\cite{trappmann}.}
\label{E1gpressure}
\end{figure}

\begin{figure}
\caption{Pressure dependence of the phase diagram with the field in the basal
plane with $(K_2+K_3)/K_1 = 0$ ($E_{2u}$ model).  The phase diagram is plotted
at pressures ($P$) of  (a) $P = 0$, (b) $P = (1/2)P_N$ (1.87 kbar), (c) $P =
P_N$ (3.7 kbar), and (d) $P = (3/2)P_N$ (5.55 kbar).  In order to obtain a
better fit in this model we have changed the values of some of our input
parameters. In these phase diagrams $a_x = a_d = 0$, $S_{c2x} =
-6.66$ T/K, $S_{c2y} = -4.07$ T/K, $T_x = 0.465$ K, and $T_y = 0.509$ K.
The data, renormalization of $H_y^*(T)$, and all other input parameters are
the same as those for Fig.\ \protect \ref{E1gpressure}.}
\label{E2upressure}
\end{figure}

\end{document}